\documentclass[preprint]{aastex} 
\newcommand{\gsim}{\mbox{$\stackrel {>}{_{\sim}}$}}  
\newcommand{\lsim}{\mbox{$\stackrel {<}{_{\sim}}$}} 
\newcommand{\thco}{$\rm^{13}CO$} 
\newcommand{\ceo}{$\rm C^{18}O$} 
\newcommand{\htwo}{$\rm H_2$} 
\newcommand{\rth}{$\rm R_{13}$} 
\newcommand{\rei}{$\rm R_{18}$} 
\newcommand{\tk}{$\rm T_K$} 
\newcommand{\nh}{$\rm n_{H_2}$} 
\newcommand{\xco}{$\rm X_{CO}$} 
 
\slugcomment{Accepted to the Astronomical Journal} 
 
\shorttitle {CO Isotopomers in NGC 6946} 
\shortauthors{Meier \& Turner}

\begin{document} 
 
\title{Dynamically Influenced Molecular Clouds in the Nucleus of NGC~6946:
Variations in the CO Isotopic Line Ratios} 
 
\author{David S. Meier\altaffilmark{1,2} \& Jean L. Turner\altaffilmark{2,3}} 
 
\altaffiltext{1}{Department of Astronomy, University 
of Illinois, 1002. W. Green St., Urbana, IL 61801  
email: meierd@astro.uiuc.edu} 
\altaffiltext{2}{Department of Physics and Astronomy, UCLA, Los Angeles, 
CA 90095-1562. 
email: turner@astro.ucla.edu} 
\altaffiltext{3}{Visiting Associate, Department of Astronomy, Caltech, 
Pasadena CA 91125}

\begin{abstract} 
 
We present high resolution ($\sim 5^{''}$) maps of the J = 1 - 0
transitions of $^{13}$CO and C$^{18}$O towards the nucleus of
NGC~6946, made with the Owens Valley Millimeter Array.  The images are
compared with existing $^{12}$CO(1-0) maps to investigate localized
changes in gas properties across the nucleus.  As compared to
$^{12}$CO, both $^{13}$CO and C$^{18}$O are more confined to the
central ring of molecular gas associated with the nuclear star
formation; that is, $^{12}$CO is stronger relative to $^{13}$CO and
C$^{18}$O away from the nucleus and along the spiral arms.  The
$^{12}$CO(1-0)/$^{13}$CO(1-0) line ratio reaches very high values of
$>$40.  We attribute the relative $^{13}$CO weakness to a rapid change
in the interstellar medium from dense star forming cores in a central
ring to diffuse, low density molecular gas in and behind the molecular
arms.  This change is abrupt, occurring in less than a beamsize (90
pc), about the size of a giant molecular cloud.  Column densities
determined from $^{13}$CO(1-0), C$^{18}$O(1-0), and 1.4 mm dust
continuum all indicate that the standard Galactic conversion factor,
X$_{CO}$, overestimates the amount of molecular gas in NGC 6946 by
factors of $\sim$3-5 towards the central ring and potentially even
more so in the diffuse gas away from the central starburst.  We
suggest that the nuclear bar acts to create coherent regions of
molecular clouds with distinct and different physical conditions. The
$^{12}$CO(1-0)/$^{13}$CO(1-0) line ratio in galactic nuclei can be a
signpost of a dynamically evolving ISM.
 
\end{abstract} 
 

\section{Introduction} 
 
Molecular clouds in galactic centers may differ significantly from the
molecular clouds we are familiar with from our own Galactic disk.
Characteristics of the nuclear regions of galaxies---strong tidal
forces, rapid rotational timescales, the propensity for molecular
cloud interactions due to confined spaces and non-circular motions,
the tendency toward extreme star formation events---can easily leave
their imprints on the properties of giant molecular clouds (GMCs), and
hence star formation.
 
Some of the first indications that GMCs in starbursts are unusual came
with the discovery of bright, high J transitions of CO and the
detection of bright HCN(1-0) lines in nearby, actively star-forming
nuclei, both indicative of warm, dense gas
\citep[eg.,][]{HTM87,NNJ89,CJTH90,THM90,SSSS91,SDR92,NJHTM92}.  In the
more extreme starbursts of distant ultraluminous infrared galaxies, CO
and HCN observations seem to indicate volume densities of greater than
$10^{3}-10^{4}$ cm$^{-3}$ over kpc scales
\citep[eg.][]{SSSS91,OKIKMI91,DS98,SDR92,GS03}. The remarkably large
amounts of high density gas in these luminous galaxies likey are the
cause of their high star formation rates.
 
Despite the certain presence of {\em high density} molecular gas,
these starbursts are often found to have low
$^{12}$CO(2-1)/$^{12}$CO(1-0) line ratios \citep[][]{RDS91,PI02} and
high $^{12}$CO(1-0)/$^{13}$CO(1-0) line ratios
\citep[][]{SI91,ABJB91,ABBJ95,CDC92,ARSS97,TO98,TOS99}, characteristic
of {\it low density} gas.  The explanation for this apparent
contradiction is not yet clear.  Explanations for the anomalously high
$^{12}$CO(1-0)/$^{13}$CO(1-0) ratios tend to fall into two main
categories: unusual physical conditions resulting in low CO opacities
or abundance anomalies.  If the CO isotopic ratios arise from diffuse,
low opacity gas, this might be an indicator of starburst feedback or
dynamical cloud dispersal, either of which could result in a diffuse
gaseous medium capable of inhibiting future star formation.  If
instead the explanation for the isotopic ratios is unusual abundances,
then the presence of high $^{12}$CO(1-0)/$^{13}$CO(1-0) might be a
useful indicator of chemical evolution of gas in the presence of star
formation.
 
In distant galaxies distinguishing between these alternatives is
difficult because of poor spatial resolution.  In nearby galaxies,
interferometric imaging at resolutions of a few arcseconds allows us
to resolve changes in physical and chemical conditions on the scale of
individual GMCs ($\sim$50 pc). Observations of the starbursts in M 82
and IC 342 have demonstrated that small-scale changes in isotopic line
ratios are present and correlate with changes in gas properties
\citep[e.g.,][]{TH92,WITHL93,NGKGW98,MTH00,WNHK01,MT01}.  In this
work, we extend such studies to the nearby, spiral nucleus with active
nuclear star formation, NGC 6946.
 
NGC 6946 is a nearby ($\sim$5.2 Mpc), fairly face-on, barred spiral
galaxy (Table \ref{galbasdat}).  One of the brightest galaxies in the
mid-infrared sky \citep[][]{RL78}, NGC~6946 is also bright in CO
\citep[][]{BSSLS85,SDINH88,WCC88,RV95,SOIS99} and radio continuum
\citep[][]{TH83}.  NGC 6946 is currently the site of a moderate
starburst that is 7 - 20 Myrs old \citep[][]{ERRL96}.
\citet[][]{BR92} find a flattened abundance gradient over the central
2.5 kpc, suggesting radial mixing of gas has occurred.
 
Here we compare interferometric $^{13}$CO(1-0) and C$^{18}$O(1-0)
observations to $^{12}$CO(1-0) with two main goals in mind: first to
obtain more reliable images of the true molecular column density in
the nucleus; and second, to correlate the observed line ratios with
other gas properties in order to determine what the isotopic ratios
tell us about the structure of the molecular ISM in NGC 6946 and
starbursts in general.
 
\section{Observations} 
 
Aperture synthesis observations of the $^{13}$CO(1-0) and
C$^{18}$O(1-0) lines in NGC 6946 were made with the Owens Valley Radio
Observatory (OVRO) Millimeter Interferometer between 1999 September 26
and 1999 November 28.  2.7 mm and 1.4 mm continuum measurements in a
separate 1 GHz bandwidth filter were also obtained.  The
interferometer consists of six 10.4 meter antennas with cryogenically
cooled SIS receivers \citep{OVRO91, OVRO94}.  System temperatures
(single sideband) ranged from 180 to 380 K at 2.7 mm.  The transitions
were observed simultaneously making use of OVRO's multiple-line
spectrometer capabilities, hence share the same instrumental
configurations, phase centers, and weather.  The instrumental
parameters are presented in Table \ref{galt2}.
 
Data were calibrated using the MMA software package.  Phase
calibration was done every 25 minutes using the calibrator 2037+511.
Absolute flux calibration was done using Neptune for primary flux
calibration and 3C273, 3C84 and 3C454.3 as secondary flux calibrators.
Due to variability of the quasars, Neptune's low elevation and
imprecise 2.7 mm brightness, the absolute flux calibration is
estimated to be good to 10 \% for the 2.6 mm data (and 20 \% for the
1.4 mm continuum map).  The dataset was reduced using the NRAO AIPS
data reduction package.  The maps are naturally-weighted and primary
beam-corrected.  In making the zeroth moment (integrated intensity)
maps, a mask was generated to help minimize the inclusion of noise
into the maps from non-signal portions of these wide lines.  The
naturally-weighted dataset was smoothed to a resolution of 10$^{''}$,
and all regions below 2$\sigma$ were blanked.  This mask was then used
on the full resolution dataset to blank out non-signal portions.  Only
emission greater than 1.2$\sigma$ was included in making the
integrated intensity maps.  In making the line ratio maps, regions
with emission below 3$\sigma$ in either map were blanked to avoid
taking ratios of low signal-to-noise emission.
 
Lack of short baselines limits our sensitivity to structures on scales
larger than $\sim$40$^{''}$ at 2.7 mm ($\sim$20$^{''}$ for the 1.4 mm
continuum map).  To estimate the amount of extended flux the
interferometer resolves out of the maps, spectra integrated over the
single-dish beamsize were compared to available single-dish spectra.
For NGC~6946, \citet{Pet01} obtain an integrated intensity of 4.8 K km
s$^{-1}$ for $^{13}$CO(1-0) (assuming $\eta_{mb}$ = 0.5 for the FCRAO
antenna).  Convolving the interferometer channel maps to the 45$^{''}$
resolution of the single-dish and sampling the map at the same
location yields a line intensity of 3.0 K km s$^{-1}$, or 63 \% of the
flux.  A similar analysis of C$^{18}$O(1-0) based on a comparison with
the 22$^{''}$ resolution spectra of \citet[][]{WBTWWD02} suggests that
$\sim$75 \% of the single-dish flux has been detected, consistent with
the $^{13}$CO(1-0) value.
 
$^{12}$CO(1-0) data has been kindly provided by S. Ishizuki
\citep[][]{SOIS99} for comparison with the isotopic line data.  While
the data were taken at the Nobeyama Radio Observatory Millimeter Array
(NRO) not OVRO, this difference should not strongly affect the maps
since the NRO array is similar in layout to OVRO with similar $(u,v)$
coverage. Also the Ishizuki $^{12}$CO(1-0) map recovers 76 \% of the
single-dish flux over the mapped region \citep[][]{SOIS99}, a
significant fraction of the total flux and very similar to that found
for the isotopomers.  Moreover, the similarity of the NRO map and the
BIMA $^{12}$CO(1-0) map, which is more sensitive to extended emission
\citep[][]{RV95,BIMASONG03}, demonstrates that the observed
distribution is not a strong function of the array used.
 
To simplify the nomenclature, we use ``CO'' to denote the main 
isotopomer, $^{12}$C$^{16}$O, $^{13}$CO to denote ``$^{13}$C$^{16}$O"; 
and C$^{18}$O to denote ``$^{12}$C$^{18}$O".  The ratio of integrated 
line intensities, $\rm \int{T_{CO(1-0)}dv}/\int{T_{^{13}CO(1-0)}}dv$ 
and $\rm \int{T_{CO(1-0)}dv}/\int{T_{C^{18}O(1-0)}}dv$ are represented 
by $\rm R_{13}$ and $\rm R_{18}$, respectively.  Since we observed the 
3 mm J=1-0 transitions, J values are omitted unless different from 
J=1-0, or where omitting results in ambiguity. 
 
\section{Results} 
\subsection{Morphology and Kinematics of the CO Emitting Gas} 
 
\begin{figure} 
\epsscale{0.6}
\plotone{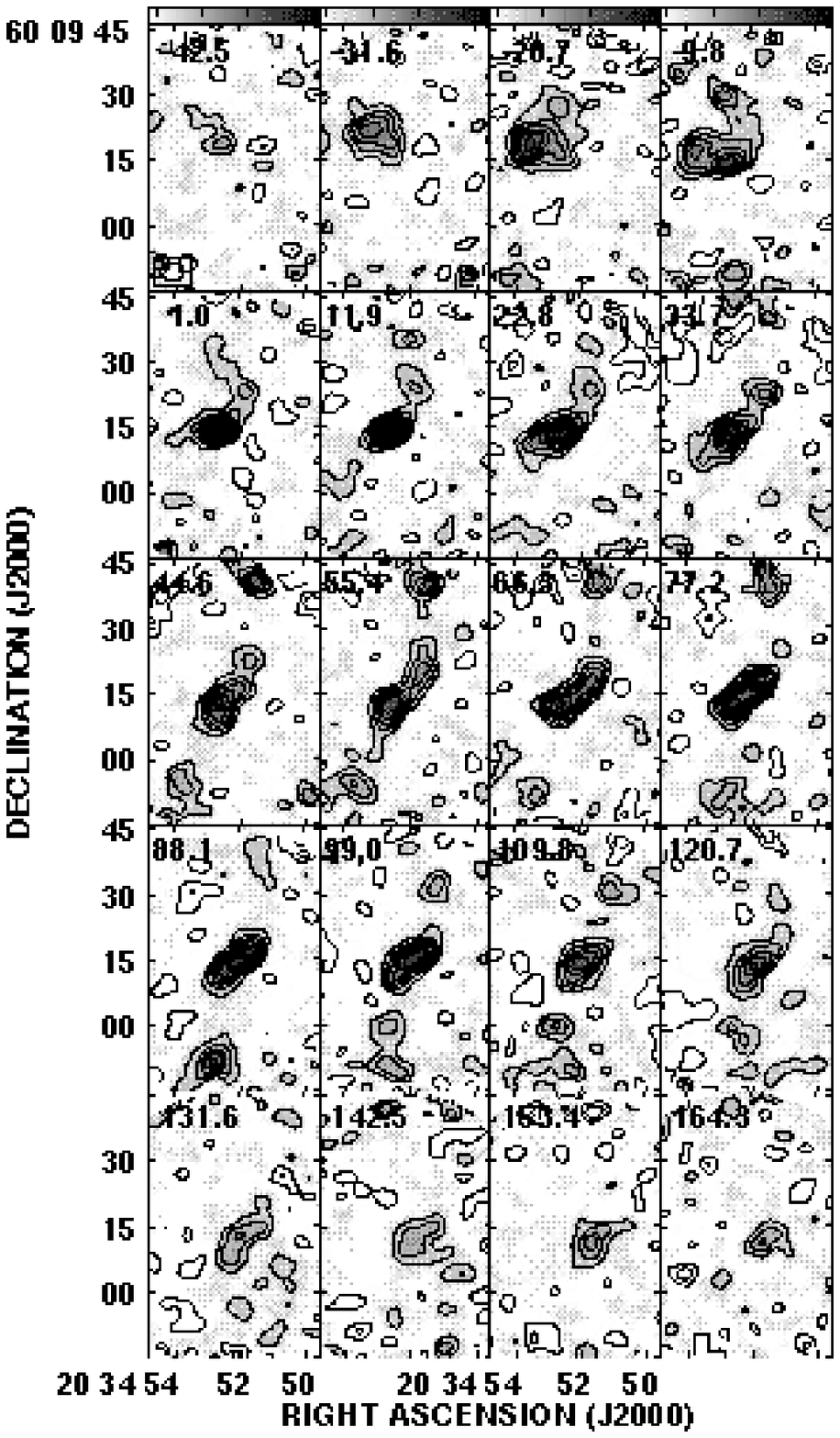}
\caption{$^{13}$CO(1-0) channel maps for the nucleus of NGC 6946.  LSR
velocities are listed at the top of each panel.  The contours are
plotted in intervals of 20 mJy bm$^{-1}$ (or 0.09 K for the
5.$^{''}$4x4.$^{''}$2 beam) corresponding to 2$\sigma$.  The beam is
plotted in the bottom left of the first panel.}
\label{galn6ch13} 
\end{figure}

\begin{figure} 
\epsscale{0.9}
\plotone{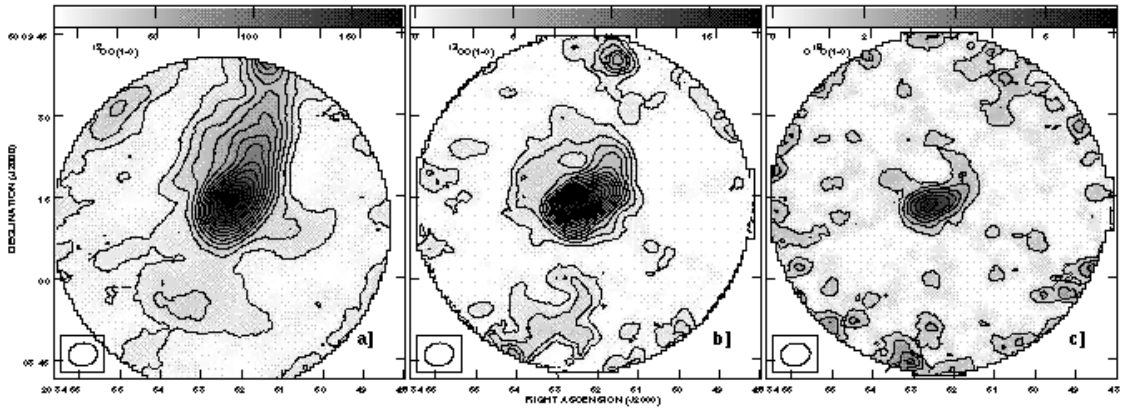}
\caption{The CO isotopomers integrated intensity maps of NGC 6946.
{\it (a)} The $^{12}$CO(1-0) integrated intensity image from
\citep[][]{SOIS99} convolved to the resolution of the $^{13}$CO(1-0)
transition (5.$^{''}$4x4.$^{''}$2).  Contours are in steps of 50.0 K
km s$^{-1}$.  {\it (b)} The $^{13}$CO(1-0) integrated intensity image.
Contours are in steps of the 2$\sigma$ value of 4.5 K km s$^{-1}$.
{\it (c)} The C$^{18}$O(1-0) integrated intensity image.  Contours are
in steps of the 2$\sigma$ value of 4.5 K km s$^{-1}$ (with the same
beamsize as in {\it (a) - (b)}).}
\label{galn6int} 
\end{figure}

The $^{13}$CO channel maps for NGC~6946 are presented in Figure
\ref{galn6ch13}.  Peak antenna temperatures obtained from the
$^{13}$CO channels are $\sim$ 0.82 K.  Emission is present in channels
from V$_{LSR}$ = -43 km s$^{-1}$ to 164 km s$^{-1}$. Redshifted
velocities are towards the southwest.  No significant differences
between CO and $^{13}$CO velocity dispersions or systemic velocities
are seen.  The morphology and kinematics (not shown) of C$^{18}$O,
though limited by faintness, are consistent with the $^{13}$CO data.
Integrated intensity maps of the three CO isotopomers are presented to
the same resolution and scale in Figure \ref{galn6int}.  Integrated
intensities of each transition are listed in Table \ref{galn6itab} for
the locations of the GMCs.
 
The CO emission within the central 1.3 kpc (50\arcsec ) of NGC~6946 is
spatially asymmetric \citep[][]{RV95, SOIS99}.  The CO integrated
intensity peaks at the starburst, and remains relatively bright along
the northern arm.  The southern arm is weaker. The optical spiral
structure of NGC~6946 also reflects this asymmetry, although on a much
larger scale, with a ``heavy'' northern arm \citep[][]{A66}.

By contrast, $^{13}$CO emission is more symmetric and spatially
confined to the center of the galaxy than is CO.  The $^{13}$CO
emission is suggestive of a central disk-like structure with the
starburst occurring at the southwestern edge of the disk.  A secondary
peak in CO at the northern end of the molecular bar is also bright in
$^{13}$CO.  Lower resolution studies of the CO showed extended
bar-like CO and hints of non-circular motion suggesting that it is a
true bar \citep[eg.][]{SDINH88, I90b}.  From their high resolution CO
map, \citet[][]{RV95} argued that it is difficult to distinguish
between a nuclear bar and a continuation of spiral arms, based on
morphology alone.  \citet[][]{ECS98} suggested the possibility of
nested bars in NGC~6946.

In Table \ref{galbasdat}, we have listed the near-IR centroid of NGC
6946 as measured from 2MASS J, H, and K images. The peak 2 cm
continuum emission (Tsai et al. 2004, in preparation) is located
$\sim$ 1\arcsec\ ENE of the near-IR center, although this emission is
extended.  The dynamical center is, within the uncertainties,
coincident with the near-IR peak, as one might expect if the near-IR
emission traces the mass.  Whereas the $^{13}$CO peak is $\sim 2^{''}$
ESE of the near-IR peak.

\begin{figure}
\epsscale{0.9}
\plotone{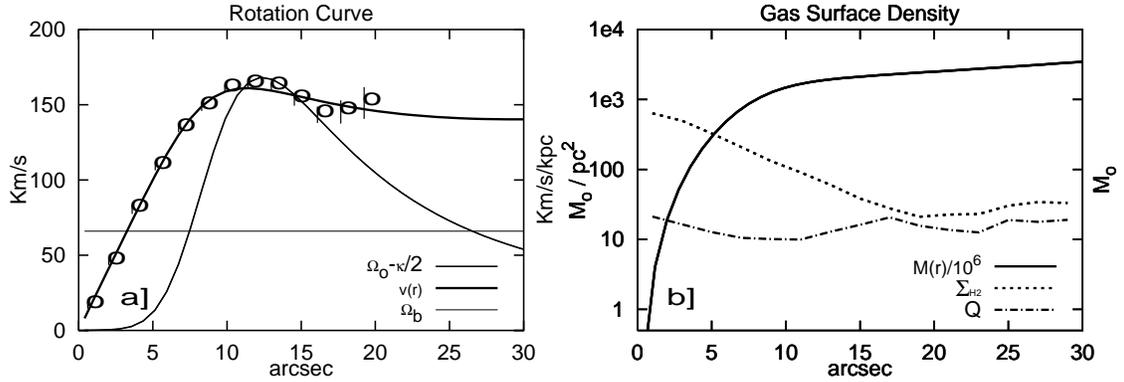}
\caption{{\it (a)} The observed and fitted rotation curve of the
central arcminute of NGC 6946.  Circles mark the measured azimuthally
averaged velocity versus radius, and short vertical bars are the
uncertainty.  The fitted model (thick solid line) consists of two mass
distributions corresponding to Brandt rotation curves (see text for
details).  The pattern speed appropriate for the nucleus, $\Omega_{p}$
= 66 km s$^{-1}$ kpc$^{-1}$ (thin solid line) is taken from
(Crosthwaite \& Turner 2004, in preparation), based on the large-scale
disk velocity field traced by CO(1-0) and HI.  $\Omega_{o}$ is the
angular velocity of the gas and $\kappa$ is the epicyclic frequency
%\citep[eg.,][]{BT87}.  Inner Lindblad resonances are present when
%$\Omega_{o}$ - $\kappa$/2 = $\Omega_{p}$.
The intersections of $\Omega_{p}$ with the $\Omega_{o}-\kappa /2$
curve (dotted line) mark the locations of the iILR ($\sim 7.5^{''}$)
and oILR ($\sim 26^{''}$) for this same pattern speed.  {\it (b)} The
observed azimuthally averaged molecular gas surface density,
$\Sigma_{H_{2}}$ (dotted line), based on the $^{13}$CO column density
derived in \S 4.2 for the central 30$^{''}$ is shown.  The cumulative
dynamical mass interior to a given radius based on the model rotation
displayed in {\it (a)} is also shown (solid line).  Given the rotation
curve in {\it (a)} and the gas surface density, we calculate and
display the Toomre $Q$ parameter (dot-dashed line) for the central
region of NGC 6946.  The surface density axis is on the lefthand side,
while the dynamical mass axis is on the righthand side in units of
$10^{6}$ M$_{\odot}$.  $Q$ uses the same axis as surface density but
is unitless.  All rotation curve data has been corrected for
inclination.}
\label{doubrot}
\end{figure}

\begin{figure} 
\epsscale{0.9}
\plotone{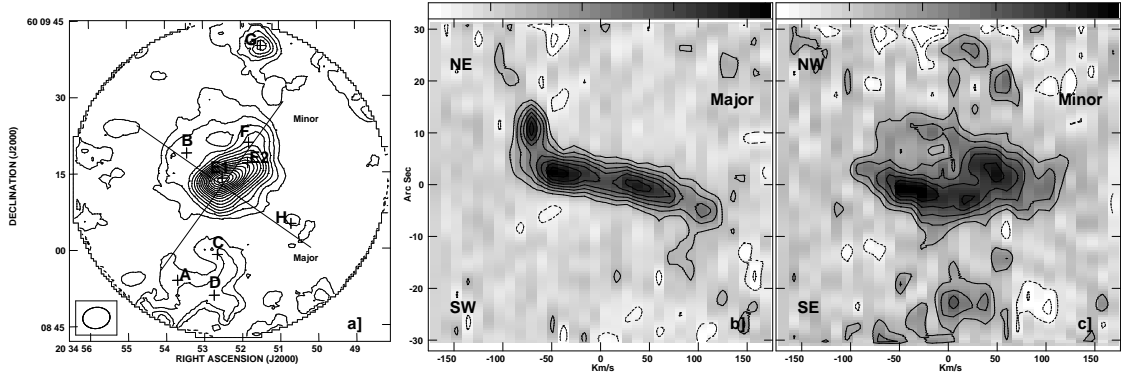}
\caption{The Position-Velocity (PV) diagrams of NGC 6946 taken from
the $^{13}$CO(1-0) data.  {\it (a)} The $^{13}$CO(1-0) integrated
intensity map of NGC 6946 contoured as in Figure \ref{galn6int}b.
Fitted GMCs are labeled on the figure as well as the locations the PV
slices have been measured at.  The PV diagram for the {\it (b)} major,
{\it (c)} minor axes.  PV diagrams are contoured in steps of 3$\sigma$}
\label{galn6pv} 
\end{figure}

To study the impact of the galactic potential on the molecular gas, we
first investigate the kinematics of the nucleus.  We use two methods
to study the velocity field in NGC 6946.  One, we fit the entire
velocity field to a disk in circular rotation and use this to
determine the azimuthally averaged rotation curve of the central
region (Figure \ref{doubrot}a).  Secondly, position-velocity (PV)
diagrams are constructed along the major and minor axes (Figure
\ref{galn6pv}).  The major axis PV diagram should essentially traced
out the same shape as the rotation curve (reflected through the
origin), but it also provides additional information on the
distribution of emission versus velocity.  Comparison of the major
axis PV diagram in Figure \ref{galn6pv}b with the rotation curve of
Figure \ref{doubrot}a demonstrates this is true.  The rotation curve
increases steeply for $\sim 7.5^{''}$ then turns over and falls for
$\sim 5^{''}$ before it reverses and continues to rise out to larger
distances.  The radius of turnover corresponds to the edge of the
central ring.  The same structure can be seen in the PV diagram.  In
Figure \ref{galn6pv}b, the two emission peaks on either side of the
center of the major axis PV diagram hint that the disk-like structure
seen in the isotopomers does not extend all the way to the center of
the galaxy and hence is ring-like (compare Figure \ref{galn6pv}b with
Figure 5 of Sakamoto et al. 1999).  The central ring-like structure
may reflect the $x_{2}$ orbits of an inner Lindblad Resonance (ILR) or
possibly gas response to the small inner bar.
 
In the case of pure circular motion the PV diagram taken along the
minor axis predicts emission only at zero velocity, since all motion
would be in the plane of the sky at these locations.  Deviations from
this imply radial motions are present.  In Figure \ref{galn6pv}c, a
large percentage of the emission over the central $5^{''}$ radius is
blueshifted on the SE side of the galaxy, and redshifted on the NW
side of the galaxy. Since the northern molecular arm is the near one
\citep[][]{ECS98,R00}, consistent with trailing spiral arms, these
shifts imply that the gas is orbiting radially inward from both sides.
Radial velocities peak at $|v_{r}|~\sim$ 80 km s$^{-1}$ at the center,
decreasing as one moves away from the center of the galaxy.  Strong
radial motion is not obvious beyond a radius of $\sim 7^{''}$.  Either
the gas inflow strengthens as the nucleus is approached, or more
likely the gas streamlines are turning progressively more radial with
decreasing galactocentric distance.

The overall kinematics of NGC 6946's nucleus can be explained either
as non-circular motions in response to a barred potential, or as a
region of enhanced mass density in the nucleus.  In the case of a
barred potential, non-circular motions ``contaminate'' the rotation
curve causing the observed rotation curve to develop breaks
\citep[eg.][]{SOIS99}.  Radial inflow will raise the inferred circular
velocities.  If the radial flow dominates over a narrow radii around
the molecular ring, the velocity peak at $\sim 7^{''}$ is explained.
Therefore the non-circular motion associated with a barred potential
provides a good explanation for both the radial motions seen in the
minor axis PV diagram and the turnover in the rotation curve.

Alternatively, the rotation curve could be due to a compact nuclear
mass distribution.  This would affect the dynamics and the location of
orbital resonances.  In Figure \ref{doubrot}a-b, the rotation curve of
NGC 6946 is shown fit with a 2-component mass density distribution:
one with a characteristic scale of 275 pc (11$^{''}$) and a peak
velocity of 175 km s$^{-1}$, roughly consistent with the size of the
secondary nuclear bar \citep[][]{ECS98}, and the other with a scale of
5 kpc (200$^{''}$) that reaches the same peak velocity.  This
2-component mass distribution matches the observed velocity profile.
If we (somewhat arbitrarily) assume the density wave pattern speed
determined for the large scale disk (Crosthwaite et. al. 2004, in
preparation) of $\omega_{p}\simeq$ 66 km s$^{-1}$ kpc$^{-1}$, is the
same for both, the ``buckle'' in the nuclear rotation curve
corresponds to an inner ILR near the ring, at a radius of 7.5$^{''}$,
and the location of the clumps at the ends of the molecular bar would
correspond to an outer ILR, at a radius of 26$^{''}$.
 
Given the strong non-circular motions across the nuclear region, we 
favor a barred potential over a central mass for explaining the gas 
morphology and kinematics. 
 
\subsection{Properties of the Nuclear GMCs in NGC 6946} 
 
Giant molecular clouds (GMCs) in NGC 6946 are identified from the
$^{13}$CO data.  The method of localizing and fitting the GMCs is the
same as given in \citet[][]{MT01}.  Results are given in Table
\ref{galgmc} and Figure \ref{galn6pv}a.  Given a distance of 5.2 Mpc
(1$\arcsec \simeq$ 25 pc), only the larger GMCs are spatially
resolvable (fitted sizes $>$1/2 the FWHM of the beamsize).  GMCs that
are off the central ring region (A, C, D, H, G) are the most isolated
and hence their fits are the most robust.  GMCs away from the central
ring generally have deconvolved sizes $\sim$50 - 100 pc in size, FWHM
linewidths of 25 - 40 km s$^{-1}$ and predicted virial masses of
$\gsim 10^{7} ~M_{\odot}$.  The central ring GMCs have implied virial
masses of $\sim 10^{8} ~M_{\odot}$; however, while NGC 6946 is not
highly inclined ($i=40^{o}$), the GMCs that are centered along the
central ring almost certainly have linewidths contaminated by the
steep rotational velocity gradient and non-circular motions present.
The derived virial masses are consistent with the enclosed dynamical
masses (\S 4.2).
 
\subsection{The High and Variable Isotopic CO Line Ratios in NGC 6946} 
 
\begin{figure} 
\epsscale{0.9}
\plotone{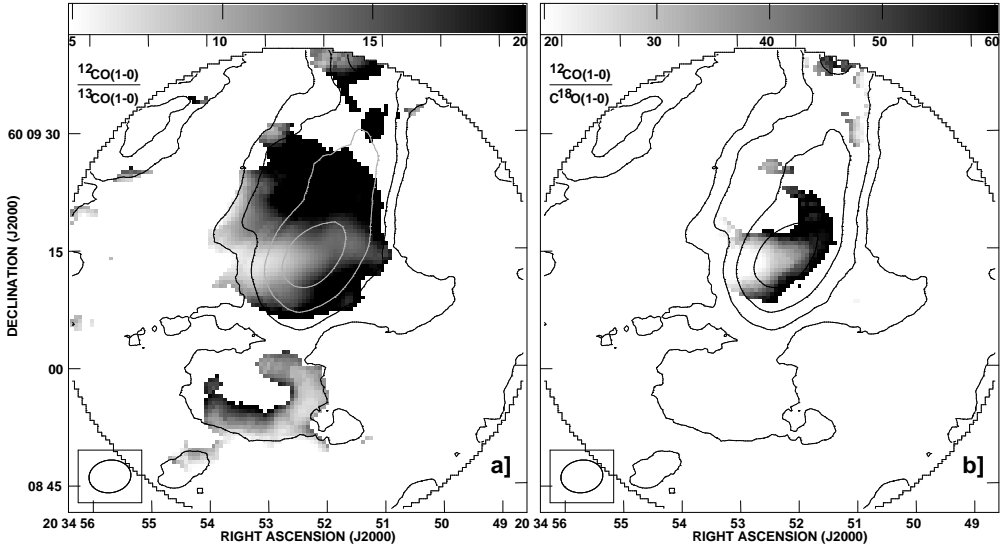}
\caption{The isotopic line ratios in NGC 6946.  {\it (a)} The
$^{12}$CO(1-0)/$^{13}$CO(1-0) line ratio ($R_{13}$).  The
$^{12}$CO(1-0) data has been kindly provided by \citet[][]{SOIS99}.
High ratios are dark.  The greyscale ranges from 5 - 20.  Contours are
$^{12}$CO(1-0) in steps of 75 K km s$^{-1}$.  The ratio map has been
blanked in regions where either map is less than 3$\sigma$ in
intensity.  In almost all cases, the transition that limits the field
is the weaker $^{13}$CO(1-0), therefore any region inside the
$^{12}$CO(1-0) contours that is blanked also has very high line
ratios.  {\it (c)} The $^{12}$CO(1-0)/C$^{18}$O(1-0) line ratio
($R_{18}$).  Contouring and greyscaling are the same as {\it (b)}
except that the greyscale ranges from 20 - 60.  Again, the
C$^{18}$O(1-0) transition is the cause of most the blanking, so
blanked regions represent high ratio regions.}
\label{galn6rat} 
\end{figure}

Figure \ref{galn6rat} (and Table \ref{galn6rattab}) shows the $R_{13}$
and $R_{18}$ isotopic line ratio maps for the nucleus of NGC 6946.
Two unusual characteristics of the line ratios are evident: there are
regions with very high values compared to the Galaxy, and the ratios
vary rapidly across the nuclear region.  Over the mapped regions,
$R_{13}$ reaches values greater than 25 along the northern arm.  By
comparison, Galactic disk molecular clouds generally have values of 6
- 7 \citep[][]{PKSW88}.  In almost all cases, $^{13}$CO (C$^{18}$O for
$R_{18}$) limits the region covered by the ratio map, which implies
that blanked regions in Figure \ref{galn6rat} have even higher
$R_{13}$, with values as high as 60 estimated for in these blanked
regions.  The lower values of $R_{13}$ are $\sim 10$ near the
starburst, more consistent with what is typically seen in starburst
nuclei \citep[eg.,][]{ABJB91,ABBJ95}.  The lowest isotopic line ratios
($R_{13}\simeq 4~-~5$) are found $\sim 5^{''}$ east of the main
starburst, extending perpendicular to the main CO arms.  Along the
northern arm gas transitions from its lowest values of $R_{13}$ to
$R_{13} \gsim$ 25 in one beamwidth ($\lsim$100 pc).  The transition
from $R_{13} \sim$ 10 regions dominating towards the starburst to
$R_{13} \gsim$ 25 is even sharper, being unresolved by our beam.
Evidently, just outside the central ring the molecular gas undergoes a
sharp change in gas physical conditions.
 
To test if the high isotopic ratio could be due to a larger amount of
flux resolved out of the $^{13}$CO transition versus the CO, we have
taken 26 \% of the total CO flux (the percentage of the flux resolved
out at CO) and distributed it uniformly over the central
30$^{''}\times 60^{''}$, and 38 \% of the $^{13}$CO and distributed it
uniformly over the same region, and then recalculated $R_{13}$.  This
approximates what we would expect to see if the flux resolved out of
each map were distributed uniformly over the largest spatial scales.
The newly calculated 2$\sigma$ lower limit to $R_{13}$ at the arm
locations become $>$47.  Therefore, the very high ratios in the arm
region cannot be due to differentially resolved-out flux..  Similar
results are seen in $R_{18}$, when taking account its more limited
coverage, providing further evidence for true changes in gas physical
conditions and not just anomalies in $^{13}$CO.  Typical values of the
$^{13}$CO/C$^{18}$O line ratio towards the high S/N regions is $\sim$
2.5.
 
Comparisons with interferometric HCN(1-0) data show that, within the
uncertainties, the morphology of the $^{13}$CO and C$^{18}$O
transitions match closely what is seen in the HCN(1-0)
\citep[][]{HB97}.  The HCN/CO ratio map of \citet[][]{HB97} correlates
well with the inner structure of $R_{13}$.  Both show a ``saddle''
shaped distributed approximately perpendicular to the CO arms
\citep[][]{HB97}.  This suggests that density changes sensitively
influence the isotopic line intensities.
 
\subsection{Millimeter Continuum from Star Forming HII Regions in  
NGC 6946} 
 
\begin{figure} 
\epsscale{0.9}
\plotone{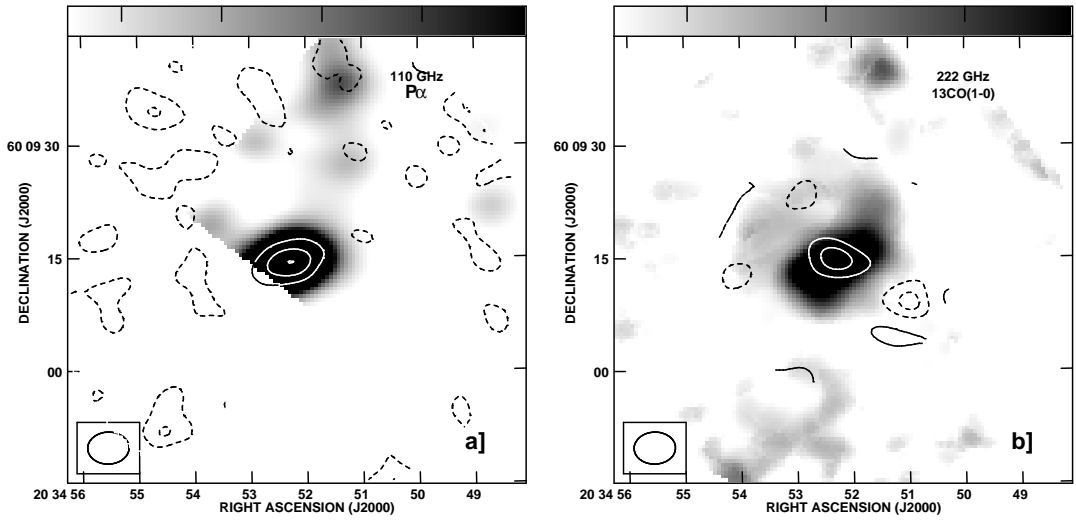}
\caption{Millimeter continuum maps of the nucleus of NGC 6946.  {\it
(a)} The 110 GHz (2.7 mm) continuum map for NGC 6946 overlaid on the
HST P$\alpha$ NICMOS image \citep[][]{HST99}, smoothed to the same
resolution.  Contours are in steps of the 2$\sigma$ value of 2.0 mJy
bm$^{-1}$.  This map has been corrected for $^{13}$CO(1-0)
contamination.  {\it (b)} The 222 GHz (1.4 mm) continuum emission from
NGC 6946 overlaid on the $^{13}$CO(1-0) greyscale.  Contours are in
steps of the 2$\sigma$ value of 7 mJy bm$^{-1}$.}
\label{galn6cont} 
\end{figure}

Displayed in Figure \ref{galn6cont} are the 2.7 mm continuum map of
NGC 6946 overlaid on the HST P$\alpha$ image \citep[][]{HST99},
convolved to the same beamsize as the OVRO maps, and the 1.4 mm
continuum map map overlaid on the $^{13}$CO(1-0) integrated intensity
map.  2.7 mm continuum is reliably detected at only one location,
1.8$^{''}$ NNW of GMC E1.  Its position is within $1^{''}$ of the 2 cm
continuum peak (Tsai et al. 2004, in preparation).  A Gaussian fit
shows that the 2.7 mm source is slightly resolved with a size of
7$^{''}\times 2^{''}$ ($PA \simeq 120^{o}$), and a total flux of
$S_{2.7mm}=8.3\pm1.0$ mJy.  The full resolution HST P$\alpha$ image,
which is sensitive to weaker star formation than the millimeter
continuum, also shows a secondary star forming peak just east of the
starburst, at the secondary HCN peak and the $R_{13}$ minima, and also
faint HII regions encircling the entire central ring
\citep[][]{HST99}.  The total flux detected at 2 cm is 23 mJy
\citep{TH83}.  The spectral index between 2 cm and 2.7 mm is
$\alpha^{2}_{2.7}\simeq -0.51\pm$0.07.  Hence, the starburst is a
mixture of bremsstrahlung and synchrotron at 2 cm, which is consistent
with the 6 - 2 cm spectral index \citep{TH83} and the distribution of
compact sources (Tsai et al. 2004, in preparation).
 
Continuum emission is also detected at 1.4 mm towards GMC E1.  The
source at 1.4 mm has a size similar to that found at 2.7 mm.  The
total flux at 1.4 mm is $21\pm 4$ mJy.  The 2.7 - 1.4 mm spectral
index implied by this flux is $\alpha^{2.7}_{1.4}=+1.4\pm0.5$.  The
rising spectral index suggests that dust emission begins to contribute
at 1.4 mm. Assuming for simplicity that all of the 2.7 mm flux is a
mixture of thermal free-free emission with a spectral index of
$\alpha$=-0.1 and dust emission with a spectral index of $\alpha$=+3.5
(slightly overestimates the free-free emission since some 2.7 mm may
still be synchrotron), we derive free-free fluxes of $\sim$ 7.0 mJy at
2.7, and $\sim 6.5$ mJy at 1.4 mm, and dust fluxes of 1.2 mJy at 2.7
mm, and 14$\pm$5 mJy at 1.4 mm. The dust flux is confined within the
$\sim 6^{''}$ associated with GMC E1.  The molecular mass implied by
the dust emission is discussed in section 4.2.
 
Using the 2.7 mm continuum to constrain the number of ionizing
photons, (and hence massive stars), a Lyman continuum rate of
N$_{Lyc}$ = 2.1$\pm0.2\times 10^{52}$ s$^{-1}$, or 2800 ``effective''
O7V stars is obtained \citep[assuming T$_{e}$ = 8000 K; eg.][]{MH67,
VGS96}.  This is a factor of two higher than is estimated from
Br$\gamma$ data by \citet{ERRL96}.  High extinction is probably the
explanation for the difference, given the $\rm H_2$ column densities
are at least $\rm \sim 1-3 \times 10^{22}~cm^{-2}$ ($\rm A_V\sim
6-17$; Table \ref{galn6col}).
 
\section{Discussion} 
 
\subsection{Physical Conditions of the Molecular Clouds in NGC 6946} 
 
\begin{figure} 
\epsscale{0.7}
\plotone{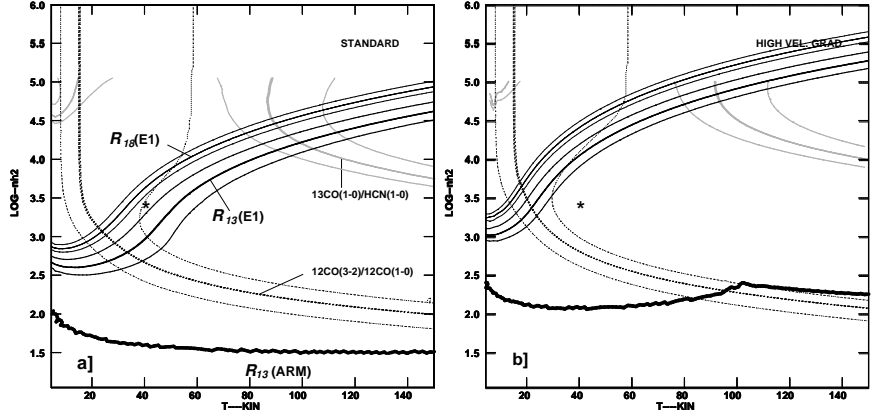}
\caption{CO LVG Models.  {\it (a)} LVG models for the $R_{13}$ and
$R_{18}$ line ratios for a ``standard'' abundance per velocity
gradient determined by $^{12}$CO/H$_{2 }$ = $8.5 \times 10^{-4}$, and
$dv/dr$ = 3.0 km s$^{-1}$ pc$^{-1}$ and [$^{12}$CO/$^{13}$CO] = 40 and
[$^{12}$CO/$^{13}$CO] = 150.  The observed values towards E1 plus
their $\pm 1 \sigma$ range of $R_{13}$ and $R_{18}$ are plotted (Table
\ref{galn6rattab}).  The fact that the acceptable parameter space
implied from $R_{13}$ is slightly lower than that implied by $R_{18}$,
suggests that [$^{13}$CO/C$^{18}$O] has been slightly overestimated.
The two curves coincide for an abundance ratio of [CO/$^{13}$CO]
$\simeq$ 60 and [CO/C$^{18}$O] $\simeq$ 150.  Also overlaid for
reference is the location where the $^{13}$CO(1-0)/HCN(1-0) line ratio
is unity assuming $X_{HCN}/dv/dr$ = $1.5\times 10^{-8}$/3 km s$^{-1}$
(see text). The single-dish $^{12}$CO(3-2)/$^{12}$CO(1-0) line ratio
is shown, along with an asterisk marking the derived kinetic
temperature and density of \citep[][]{WBTWWD02}.  Overlap of the
large-scale CO(3-2)/CO(1-0) ratio with $R_{13}(E1)$ gives an estimate
of the large-scale average excitation.  On the other hand, the overlap
between $R_{13}(E1)$ and $^{13}$CO(1-0)/HCN(1-0) gives an indication
of the local ($\sim 5^{''}$) physical conditions toward E1.  The
$R_{13}$ = 30 contour is also displayed to show the constraint on the
parameter space suitable for the molecular arm regions (labeled
$R_{13}$(ARM)).  Since along the arms, $R_{13} \gsim$ 30, acceptable
densities are $n_{H_{2}} \lsim 10^{1.8}$.  {\it (b)} Same plot as in
{\it (a)} except for a velocity gradient (abundance) four times larger
(smaller).  This gives an indication of the sensitivity of the models
to changes in abundance per velocity gradient, and may represent
models more suitable to the large linewidth central molecular ring.}
\label{lvgfig} 
\end{figure}

To relate the observed intensities and line ratios to physical
properties of the clouds, $n_{H_{2}}$ and T$_{k}$, a sample of Large
Velocity Gradient (LVG) radiative transfer models were run
\citep*[eg.,][]{GK74, SS74, DCD75}.  For very opaque lines, LVG models
can be unreliable, but for the lower opacity $^{13}$CO and C$^{18}$O
transitions LVG models are expected to yield reasonable results.  The
LVG model used is the same basic model discussed in \citet{MTH00},
expanded to cover $n_{H_{2}}$ = $10^{1}-10^{6}~cm^{-3}$, T$_{k}$ = 1.5
- 150 K, and $X_{CO}$/$dv/dr$ = 10$^{-3}-10^{-7.7}$.  A [CO/H$_{2}$]
abundance of $8.5\times 10^{-5}$ \citep[][]{FLW82} and a velocity
gradient of $3 ~km~s^{-1}~pc^{-1}$ was adopted as the standard case.
A wide range of $^{13}$CO and C$^{18}$O relative abundances and
velocity gradients were tested.  Figure \ref{lvgfig}a displays the
results assuming [CO/$^{13}$CO] $\simeq$ 40 and [CO/C$^{18}$O] $\simeq
~ 150$, typical values for starburst nuclei \citep[][]{HM93,WR94,W99}.
Only ratios are used to constrain parameter space, so the derived
solutions are independent of filling factors, as long as they are the
same for both transitions.
 
In addition to the CO LVG models, a set of HCN LVG models were run.
The first 12 J transitions were included since collision coefficients
are available only for those \citep[][]{GT74}.  We do not extend the
models to densities larger than $10^{5}~cm^{-3}$ to avoid significant
populations in J$_{max}$; our derived densities are lower than this
anyway.  An HCN abundance, [HCN/H$_{2}$] = $1.5 \times 10^{-8}$,
consistent with galactic HCN abundances \citep[eg.,][]{IGH87,PJBH98},
and a velocity gradient of $3 ~km~s^{-1}~pc^{-1}$ was assumed.  From
the observed ratios R$_{13}$, R$_{18}$ and $^{13}$CO/HCN, physical
conditions for the starburst region, E1, were estimated.
 
Since both the $^{13}$CO and C$^{18}$O lines are optically thin (\S
4.2) $R_{13}$ and $R_{18}$ should provide similar constraints on gas
density, \nh, and kinetic temperature, \tk, with any differences
ascribed to inaccuracies in the adopted [$^{13}$CO/C$^{18}$O]
abundance ratio.  Both $R_{13}$ and $R_{18}$ predict $n_{H_{2}}\sim
10^{2.5-3}$ cm$^{-2}$ and T$_{k}~\sim$ 10 - 40 K ranging to
$n_{H_{2}}\sim 10^{4.5}$ cm$^{-2}$ and T$_{k}~\sim$ 100 - 150 K.
\rth\ predicts a slightly lower range in \nh, \tk\ than does \rei.
This suggests that the adopted [$^{13}$CO/C$^{18}$O] abundance ratio
is slightly overestimated.  The solution ranges agree if the
[$^{13}$CO/C$^{18}$O] abundance ratio is lowered from 3.75 (150/40) to
$\sim$2.5 (150/60).
 
Further constraints on \nh\ and \tk\ can be obtained by adding the
$^{13}$CO/HCN line ratio, from HCN observations of \citet[][]{HB97} at
a nearly identical beamsize.  Towards GMC E1, $^{13}$CO/HCN $\simeq
1.0\pm 0.2$.  The similarity in intensity of the $^{13}$CO and HCN
lines immediately argue for high densities ($n_{H_{2}}\gsim 10^{4.5}$)
as required to thermalize HCN.  The best fit, including HCN, is
$n_{H_{2}}\sim 10^{4.2}$ cm$^{-3}$ and T$_{k}~\sim$ 90 K.  The
single-dish value of a characteristic $\Delta$J line ratio,
CO(3-2)/CO(1-0) is 0.71$\pm$0.2 \citep[][]{WBTWWD02}.  Using this line
ratio with $R_{13}$ or $R_{18}$ favors $n_{H_{2}}\sim 10^{3}$
cm$^{-3}$ and T$_{k}~\sim$ 30 K, consistent with the $n_{H_{2}}\sim
10^{3.4}$ cm$^{-3}$, T$_{k}~\sim$ 40 K obtained by
\citet[][]{WBTWWD02}.  The fact that the high resolution
interferometer data towards the starburst favors warmer, denser
solutions than those obtained from single-dish data suggests that the
molecular clouds are cooler and less dense outside the central
starbursting ring and that this lower excitation, diffuse molecular
component dilutes the single-dish ratios.
 
In the case of the high $R_{13}$ ratio arm regions, for [CO/$^{13}$CO]
= 40 and $R_{13} > 40$, no portion of parameter space in the standard
model is acceptable.  Assuming LVG is applicable, we are forced to
conclude that the $^{13}$CO relative abundance is lower than the
assumed value of [CO/$^{13}$CO] = 40, and densities are $n_{H_{2}}
\lsim 10^{2}$ cm$^{-3}$.  At these low densities $^{13}$CO emission is
strongly subthermal, and $\rm R_{13}$ and R$_{18}$ are approximately
independent of \tk.  These constraints can be loosened somewhat by
raising $X_{CO}/dv/dr$ significantly (Figure \ref{lvgfig}b), but it is
unclear how the velocity gradients could be stronger along the arms
than in the high linewidth, central ring region (Table \ref{galgmc}).
Single-dish observations of the higher J transitions CO(2-1)/CO(1-0),
CO(3-2)/CO(2-1) and CO(4-3)/CO(3-2) line ratios all decrease up the
northern arm \citep[][]{WJBIMB93,NDBW99,IB01,WBTWWD02}.  While not
extremely low, these ratios do suggest that even the higher opacity
(less subthermal) CO transitions may not be highly excited.
 
In summary, the LVG solutions for \nh\ and \tk\ based on single-dish
ratios ($\sim 22^{''}$ beam) are lower than our interferometer
solutions because the large single-dish beamsize averages the low
density, lower temperature spiral arm gas with the warmer, denser
molecular gas associated with the starburst.  The bright, compact
starburst gas dominates the interferometer maps. On the basis of the
incompatibility of the line ratios with the CO line intensity, it has
been argued that the nuclear ISM in NGC 6946 requires at least two
distinct components with distinct physical conditions
\citep[][]{Pet01,IB01}.  Our interferometer images confirms this
conclusion, and resolves the spatial distribution of each component.
 
\subsection{CO as a Tracer of Molecular Gas Mass: The Conversion Factor  
in the Nucleus of NGC~6946} 
 
One might expect that the unusual and varying cloud conditions in NGC
6946's nucleus will have an affect the CO conversion factor, \xco,
which is based on Galactic clouds that have relatively uniform
properties. Therefore we consider in detail an estimate of the
conversion factor suitable to the nucleus of NGC~6946.  To obtain a
robust estimate of the amount of molecular gas, column densities
derived from the optically thick CO integrated intensity and \xco\ are
compared with four independent estimates of the column density: from
the optically thin $^{13}$CO and C$^{18}$O lines, from the virial
theorem, and from dust.
 
Table \ref{galn6col} records the column densities derived from CO and
a Galactic conversion factor of $X_{CO} ~= ~2.0\times 10^{20}$
cm$^{-2}$ $(K~km~s^{-1})^{-1}$ \citep[][]{Set88,Het97}. Column
densities are derived from the isotopomers using:
\begin{equation} 
N(H_{2})_{^{i}CO}=2.42\times 10^{14}~cm^{-2}~ {[~H_{2}~] \over
[^{i}CO]}~{e^{{^{i}E_{u} \over T_{ex}}} \over (e^{{^{i}E_{u} \over
T_{ex}}} - 1)}~ I_{^{i}CO}~(K~ km~ s^{-1}),
\end{equation} 
where $^{i}E_{u}$ is 5.29 K (5.27 K) for $^{13}$CO (C$^{18}$O), and
abundances listed in Table \ref{galn6col}.  Excitation temperatures of
$^{13}$T$_{ex}$ = 20 K and $^{18}$T$_{ex}$ = 10 K were adopted, taking
into account the fact that the isotopomers are likely increasingly
subthermal at the relevant densities (\S 4.1).  Derived column
densities vary approximately linearly with the chosen T$_{ex}$, so
uncertainties in $\rm N_{H_2}$ scale with T$_{ex}$.  Non-LTE effects
such as temperature gradients may explain some portion of the high
observed $R_{13}$ ratio \citep[eg.,][]{MTH00}, although we doubt the
opacities are high enough here (see below).  Taking into account
uncertainties in T$_{ex}$ and abundances, we expect that column
densities of N$_{H_{2}}$ derived from isotopomers are accurate to
about a factor of two.
 
For molecular clouds with density profiles given by $\rho \propto
R^{-1}$, the virial mass is: $\rm M_{v}$ = 189 $(\Delta v_{1/2}
~km~s^{-1})^{2}$ $(R/pc),$ \citep*[eg.,][]{MRW88}, where R is the
radius of the cloud, and $\Delta v_{1/2}$ is the FWHM of the linewidth
(following Meier \& Turner 2001).  The radius of the clouds is assumed
to be 0.7$\sqrt{ab}$, where $a$ and $b$ are the FWHM fitted sizes of
the major and minor axes.  Virial cloud masses are displayed in Table
6.  Virial masses are always much larger than those found by any other
method, even for GMCs far away from the central high linewidth
molecular ring.  In general, the virial estimates of column densities
are much greater even than the column densities estimated from CO
\citep[e.g.,][]{MT01}. This confirms that GMC linewidths are not
virial across the nucleus and are dominated by rotational and
non-circular motion.
 
For clouds that are entirely molecular and have a constant gas-to-dust
ratio of 100 by mass, the gas mass is related to the 1.4 mm dust
continuum flux by \citep[eg.,][]{H83}:
\begin{equation} 
M_{gas}(1.4~ mm)~=~310~ M_{\odot} 
\left(\frac{S_{1.4mm}}{mJy} \right) 
\left(\frac{D}{Mpc} \right)^{2} 
\left(\frac{\kappa_{\nu}}{cm^{2}~g^{-1}} \right)^{-1} 
\left(e^{\frac{10.56}{T_{d}}}-1\right), 
\end{equation} 
where $\kappa_{\nu}$ is the dust absorption coefficient at this
frequency, $S_{1mm}$ is the 1.4 mm flux, D is the distance and T$_{d}$
is the dust temperature. T$_{d}$ = 40 K is assumed for the nucleus of
NGC 6946 \citep[][]{E91}, although the T$_{d}$ suitable for the 1.4 mm
emission could be lower.  The dust opacity, $\kappa_{\nu}$, at 220 GHz
is taken to be $3 \times 10^{-3}~\rm cm^{2}~g^{-1}$, but is uncertain
to a factor of four \citep{PHBSRF94}.  Dust masses determined from
millimeter continuum are only linearly dependent on T$_{d}$, so dust
opacity dominates the uncertainty in the mass.  The derived total
molecular gas mass of GMC E1 based on its dust emission, is
M$_{d}(M_{\odot})$ $\simeq$ $1.2\pm0.4 \times 10^{7}~M_{\odot}$.  The
mass is consistent with the $^{13}$CO mass.  If one instead uses the
dust extinction towards the starburst derived from the NIR, A$_{V}$
$\gsim$ 10.4 \citep[][]{ERRL96}, assumes the starburst is in the
center of the GMC, and applies the Galactic conversion between
N(H$_{2}$) and A$_{v}$ \citep[][]{BSD78}, a column density of
N$_{H_{2}}$ $\sim 2 \times 10^{22}$ cm$^{-2}$ is estimated, within a
factor of 50 \% of the value obtained from $^{13}$CO.
 
Column densities obtained from \thco\ and \ceo\ are lower than what is
implied by CO: values estimated from $^{13}$CO are low by a factor of
5$\pm$2, and those from C$^{18}$O are low by a factor of 3$\pm$2.  The
agreement between the $^{13}$CO and C$^{18}$O masses is generally good
in the high signal-to-noise regions, with the exception of E2.  The
slight difference between the magnitude of $X_{CO}$ between $^{13}$CO
and C$^{18}$O can be explained in one of two ways.  Either $^{13}$CO
is optically thick, with opacities, $^{13}\tau~\sim$ 5/3, or that the
relative isotopic abundances that we have assumed are incorrect (\S
4.1).  It is unlikely that the isotopomers are optically thick: even
at E1, an opacity of $^{13}\tau~\sim$ 0.1 is estimated given the
observed linewidth; along the northern arm, even CO may not be very
optically thick in this region (assuming CO is optically thin one
obtains column densities less than a factor of two higher than the
$^{13}$CO value).  The more likely explanation is that the relative
abundance ratio is incorrect.  Given that $^{13}$CO and C$^{18}$O are
optically thin, the $^{13}$CO/C$^{18}$O line ratio should be close to
the [$^{13}$CO/C$^{18}$O] abundance ratio.  Adopting an average
$^{13}$CO/C$^{18}$O line ratios of $\sim$2.5, and a [CO/C$^{18}$O] =
150, the favored [CO/$^{13}$CO] abundance is $\simeq$ 60.
 
The CO/$^{13}$CO abundance ratio in NGC 6946 would have to be $\sim$
200, and CO/C$^{18}$O $\sim$ 750 to make the isotopomer masses
consistent with values derived from the Galactic X$_{CO}$.  While it
is possible that towards the starburst the excitation temperature is
underestimated, a very high $\rm T_{ex}$ of 40 - 80 K is required to
match the Galactic X$_{CO}$.  In both cases, these are much higher
$\sim$50 pc averages than seen anywhere in the Galaxy.  Moreover, the
dust masses agree with the values obtained from the isotopomers toward
the detected GMC, E1, also favoring a lower conversion factor.
\citet[][]{R00} has undertaken a detailed, multicolor NIR extinction
study of several galaxies including NGC 6946 and derives conversion
factors of $^{6946}X_{CO}$ $\simeq$ $0.2 \times 10^{20}$ $cm^{-2}~(K
~km~s^{-1})^{-1}$ at the northern arm and $0.5 \times 10^{20}$
$cm^{-2}~(K ~km~s^{-1})^{-1}$ for the central ring area.  The analysis
here confirms the low X$_{CO}$ found by the \citet[][]{R00} study, and
suggests that the northern arm, in particular, is relatively
overluminous in CO for its column density and mass.
 
In summary, though the uncertainties are still large due to the
weakness of the isotope emission in NGC 6946, $^{6946}X_{CO}$ $\simeq$
$0.5 \times 10^{20}$ $cm^{-2}~(K ~km~s^{-1})^{-1}$ for the central 300
pc, possibly dropping another factor of a few away from the central
region along the northern arm.  This is at least a factor of four
smaller than $\rm X_{CO}$ in the Galaxy, and is consistent with other
findings of low $\rm X_{CO}$ in the centers of our own and other
galaxies \citep[eg.,][]{DHWM98,R00,MT01}.
 
\subsection{The Mass of \htwo\ Gas and Efficiency of the Starburst  
in NGC 6946} 
 
Over the central 130 pc (6$^{''}$) of the NGC 6946 the total molecular
mass estimated from $^{13}$CO is $2.8 \times 10^{7}$ M$_{\odot}$,
lower than previous estimates using the Galactic conversion factor.
The dynamical mass over the same region (Figure \ref{doubrot}b) is
M$_{dyn} ~=~ 5.1 \times 10^{8}$ M$_{\odot}$.  This implies a molecular
mass fraction of $\sim$ 5 \%. The kinematics of the molecular gas in
the nuclear region must be governed by the underlying stellar mass
distribution.
 
We can estimate the current star formation efficiency, SFE, based on
the gas mass and a stellar mass from the young star formation
associated with the millimeter continuum emission.  Adopting a ZAMS
Salpeter IMF with an upper mass cutoff of 100 M$_{\odot}$ and a lower
mass cutoff of 1 M$_{\odot}$, the total mass of newly formed stars
derived from the observed N$_{Lyc}$ rate (\S 3.4) is M$_{*}~ \simeq ~
4\times 10^{5}$ M$_{\odot}$, and up to $10^6 ~\rm M_\odot$ if the IMF
goes down to M stars.  Therefore, the SFE = M$_{*}$/(M$_{*}$ +
M$_{gas}$) is $\lsim$ 3\%.  This is typical of what is seen in
Galactic disk GMCs on similar sizescales
\citep[eg.,][]{MDTCSDH86,LBT89,KK03}.  Evidently, star formation in
the nucleus of NGC~6946 is not exceptional in efficiency, only being
as intense as it is because of the large amount molecular gas present.
 
Often star formation is said to result from gravitational disk
instabilities, typically characterized by the Toomre $Q$ parameter
\cite[eg.][]{K89}.  In the nucleus of NGC 6946, $Q$ is always much
larger than one even if the Galactic conversion factor is assumed, due
primarily to the large linewidth (Figure \ref{doubrot}d). Star
formation traced by the 2.7 mm continuum peaks towards the gas column
density peaks traced by $^{13}$CO, but the 2.7 mm continuum only
samples the brightest HII regions.  However, star formation traced by
P$\alpha$ also closely correlates with the $^{13}$CO, more so than
with CO.  This argues that star formation is more closely correlated
with gas surface density, than the ratio of gas surface density to
critical surface density ($Q$). In reality, it is likely that the star
formation actually correlates with {\it density} than surface density
(differs just by a multiplicative factor if the line of sight
dimension is approximately constant), since the isotopomer peaks
represent density peaks (\S4.1) and, in general, far-IR fluxes are
more tightly correlated with HCN than they are with CO
\citep[][]{GS03}.
 
\subsection{Causes of High R$_{13}$ in NGC 6946: Low Density and 
CO and C$^{18}$O Abundance Enhancements} 
 
A number of possible explanations for the simultaneously very bright
CO and weak $^{13}$CO has been suggested in the literature, primarily
in the context of strong starbursts associated with distant mergers.
Two main classes of possibilities are envisioned for generating high
$R_{13}$ ratios in NGC~6946.  A first class of possibilities requires
unusual isotopic abundances.  This might occur through the inflow of
metal poor disk gas from the outer disk, isotope-selective
fractionation or photodissociation, or chemical evolution due to
pollution from in situ nuclear star formation
\citep[][]{CDC92,HM93,TO98,TOS99}.  The second class of possibilities
involve changes in physical conditions of the molecular gas which
lower the CO opacity.  This could be caused by, unusually high average
gas temperatures, unusually low average gas densities, or unusually
high cloud velocity dispersions \citep[][]{ABJB91,ABBJ95,GH01,Pet01}.
Previous work has been limited by lack of spatial resolution, and it
is clear from these ratio maps that the cloud properties vary rapidly
across the nucleus.  The high spatial resolution ($\lsim$100 pc) of
these images \citep[and also IC 342:][]{MTH00,MT01} and this nearby
galaxy allow for a clearer understanding of what molecular traits are
most closely connected to the high $R_{13}$ gas.

\subsubsection{High $R_{13}$ from Opacity Effects}
 
\begin{figure} 
\epsscale{0.9}
\plotone{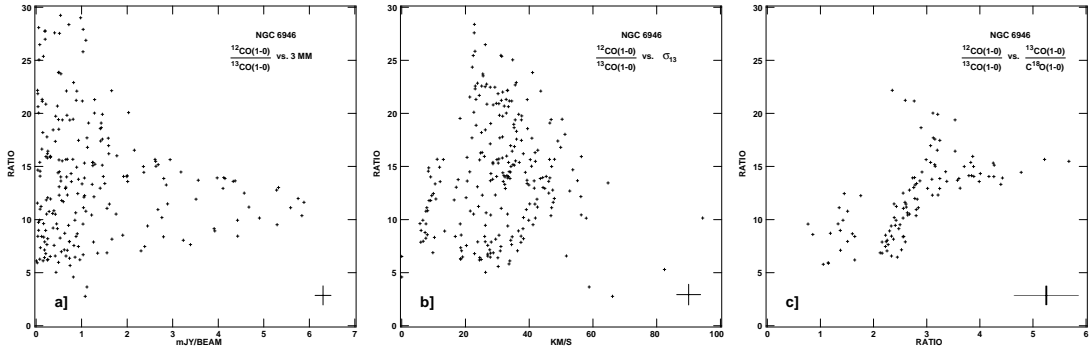}
\caption{{\it (a)} R$_{13}$ vs. 3 mm continuum sampled at 1$^{''}$
intervals.  The cross at the lower righthand corner of the figure
displays representative errorbars for each point.  {\it (b)} Same as
{\it (a)} except for R$_{13}$ vs. $^{13}$CO(1-0) velocity dispersion.
{\it (c)} Same as {\it a)} except for R$_{13}$
vs. $^{13}$CO(1-0)/C$^{18}$O(1-0).  Note: Since the sampling rate is
finer than the beamsize not all individual points are independent.
Data is only displayed for the high S/N regions not blanked in Figure
\ref{galn6rat}.}
\label{galratvs3mm} 
\end{figure}

We first discuss models which account for the anomalously high \rth\
with changes in line opacity due to variations in physical conditions,
such as \nh, \tk, or velocity dispersion.  The most direct way to
raise \rth\ (or \rei ) is to decrease the opacity of each line.  The
first way to do this is to raise \tk.  As \tk\ increases the optically
thinner low J transitions of $^{13}$CO (and C$^{18}$O) depopulate,
lowering their opacities and brightness temperatures, while the
brightness temperature of the optically thick CO line increases.  So
long as the $^{13}$CO opacity is not greater than unity, $R_{13}$ will
increase as \tk\ increases.  In NGC 6946 this is probably not the
primary mechanism for enhanced $R_{13}$ since the regions with the
highest $R_{13}$ are not found near the starforming regions, where the
gas should be warmest (Figure \ref{galratvs3mm}a).  However, changes
in \tk\ cannot be ruled out completely until the gas temperature
distribution is determined from high-resolution, multi-line analysis.
 
A second way of lowering the opacity of the molecular gas is
dynamical.  Broadening the linewidth of the molecular gas spreads the
total column density over a larger linewidth, reducing the opacity at
a given velocity, hence raising $R_{13}$ \citep[][]{ABBJ95}.  It is
clear from Figure \ref{galratvs3mm}b that there is no correlation
between $R_{13}$ and $\Delta v$ locally in NGC 6946.  In fact, the
locations of the highest linewidth are found towards the central ring,
but these areas tend to have the lowest $R_{13}$ (Table
\ref{galn6rattab}).  Moreover, if in general changes in linewidth
dominated, one would expect a correlation between $R_{13}$ and galaxy
inclination which is not seen \citep[][]{SI91,Pet01}. It is unlikely,
therefore, that broadened linewidths are the reason for the high
$R_{13}$.

Alternatively, $R_{13}$ can be increased by reducing the gas density.
Due to radiative trapping, the critical density of CO is much lower
than for \thco\ or \ceo.  The critical density for the $^{13}$CO line
is $n_{cr}~\simeq ~ 2\times 10^{3}$ cm$^{-3}$, compared to $2\times
10^{3}/\tau_{12CO}$ cm$^{-3}$, or typically $\sim$ 300 cm$^{-3}$, for
CO.  Line intensities drop rapidly when the gas density is below the
critical density.  At densities of a few hundred cm$^{-3}$ CO is still
bright but $^{13}$CO and C$^{18}$O are subthermally excited and faint.
Moreover, at such low densities intensities of the dense gas tracers
such as HCN ($n_{cr}~\sim ~ 10^{5-6}/\tau_{HCN}$ cm$^{-3}$) decrease
dramatically.  This scenario explains well the overall structure of
$R_{13}$ in NGC 6946 and in IC 342 \citep[][]{MTH00, MT01}.  In fact,
in galaxies with gas densities of $\sim 10^{3}$ cm$^{-3}$, the
isotopomers become powerful density probes (note that the $R_{13}$
contours are approximately horizontal in Figure \ref{lvgfig} at
densities below $\sim 10^{4}$ cm$^{-3}$).  This also provides a
natural explanation for the ``underluminosity'' of $^{13}$CO versus
``overluminous'' CO typically seen in distant starbursts
\citep[][]{TO98}.
 
\subsubsection{High $R_{13}$ from Abundance Variations} 
 
The second major class of models that produce enhanced \rth\ are
caused by changing molecular abundances.  The CO opacity can be
lowered by decreasing the overall [CO/H$_{2}$] abundance. This is
unlikely given the active star formation history and high metallicity
of the nucleus \citep[$3~Z_{\odot}$;][]{BR92}.  Moreover, since CO is
optically thick [CO/H$_{2}$] must be decreased by a large amount
before a strong effect is seen in $R_{13}$.  Decreasing the isotopic
abundances equate directly to changes in $R_{i}$ due to their low
opacity, so changes in the relative isotopic abundances have a much
more direct impact on $R_{i}$.  Several methods of changing the
isotopic abundances are considered in turn.
 
One way to raise the [CO/$^{13}$CO] abundance ratio is from the inflow
of $^{13}$CO-poor molecular gas from the galactic disk.  In the
Galaxy, the [CO/$^{13}$CO] abundance ratio increases with
galactocentric distance \citep[eg.][]{WR94,W99}. Driving \thco-poor
gas into the nucleus could raise the average [CO/$^{13}$CO].  NGC 6946
is a barred galaxy, and bars promote radial gas flow, which also tends
to smooth out radial abundance variations \citep[eg.][]{AELP81,FBK94}.
Flat abundance profiles over the central 2.5 kpc of NGC 6946 are
already established for N and O \citep[][]{BR92}.  Therefore, it is
possible that the average [CO/$^{13}$CO] abundance ratio over the
region is slightly high compared to Galactic ratios (as suggested by
the LVG analysis).  However, the abundance ratio should be well mixed
over the whole nuclear region, while in fact, in both NGC 6946 and IC
342, nuclear $R_{13}$ values are significantly larger than those found
in the disk \citep[][]{Pet01}.  So even if gas inflow is important it
still requires that the molecular gas undergo a change in physical
conditions as it enters the nuclear region in order to explain the
increase in $R_{13}$.  Additionally, [CO/C$^{18}$O] increases with
Galactocentric distance with a slope steeper than [CO/$^{13}$CO].
Hence, if inflow alone was the determining factor, the
$^{13}$CO/C$^{18}$O ratio would be expected to be driven to even
higher values, more typical of galactic disks, $\gsim$5.  In fact, the
opposite is seen.
 
Isotopic abundances can also be changed by chemical fractionation
and/or isotope-selective photodissociation.  Chemical fractionation
imparts a chemical formation bias towards $^{13}$CO over CO when gas
is cold and ion-molecule chemistry dominates \citep[significantly
below 35 K; eg.][]{LGCW80,LGFA84}, while isotope-selective
photodissociation lowers the isotopomer abundances relative to the
main isotopomer because of decreased self-shielding
\citep[eg.][]{VB88}.  It is questionable whether chemical
fractionation is relevant here since the molecular gas is mostly 35 K
or warmer (\S 4.1).  Moreover, $R_{13}$ does not correlate with star
formation (Figure \ref{galratvs3mm}a).  The locations of the highest
$R_{13}$ are well away from the starburst site, where the gas is
presumably cooler.  Likewise for isotope-selective photodissociation.
We dismiss both possibilities as controlling factors of the global
$R_{13}$ morphology.  But we note that if the low $R_{13}$ and
$R_{18}$ seen towards the eastern edge of the central ring are typical
of physical conditions in the central ring in the absence of strong
star formation then the slightly higher ratios, observed locally
towards the starburst western edge of the ring, could result from
localized selective photodissociation of $^{13}$CO and C$^{18}$O.
 
A final possibility for \rth\ enhancement due to variable molecular
abundances is enrichment due to ejecta from massive stars.  Though
still somewhat uncertain, massive stars are predicted to pollute the
nuclear ISM with $^{12}$C and potentially $^{18}$O, primarily early on
in a starburst, with $^{13}$C following later as intermediate mass
stars evolve off the main sequence
\citep[eg.,][]{WM92,HM93,PAA96,RM03}.  Enhancement of $^{12}$C and
$^{18}$O at the expense of $^{13}$C in young systems is expected if
the starburst IMF is biased to massive stars \citep[as suggested for
NGC 6946,][]{ERRL96}.  If the newly generated $^{12}$C and $^{18}$O is
incorporated back into CO rapidly then CO and C$^{18}$O will be
enhanced relative to $^{13}$CO, raising $R_{13}$, leaving R$_{18}$
relatively unaffected, and lowering the $^{13}$CO/C$^{18}$O ratio.  A
combination of low opacity in the isotopomers and a relatively low
$^{13}$CO/C$^{18}$O ratio implies that [$^{13}$CO/C$^{18}$O] $\sim$ 2
- 3 in the nucleus of NGC 6946.  As discussed in \S 4.1, this provides
some evidence that more applicable isotopic abundances are
[CO/$^{13}$CO] $\sim$ 60 and [CO/C$^{18}$O] $\sim$ 150, and
[$^{13}$CO/C$^{18}$O] $\sim$ 2.5.  The [$^{13}$CO/C$^{18}$O] abundance
ratio is significantly lower than the solar value of $\sim$6, the
Galactic center value of $\sim$10 and typical nearby galaxy nuclear
values of 4.  Thus it is possible that the somewhat low
[$^{13}$CO/C$^{18}$O] value combined with the absence of a tight
spatial correlation with the starburst indicates that the entire
nuclear region is modestly enriched in CO and C$^{18}$O as compared to
$^{13}$CO.  Orbital timescales are 8 Myr for the central ring and 28
Myr at the edge of the field of view, much shorter than the lifetime
of intermediate mass stars but comparable to the estimated starburst
age of 7 - 20 Myr \citep[][]{ERRL96}.  If the ejecta can immediately
return to the ambient ISM then interstellar enrichments probably has
had time to spatially mix, providing an explanation for the globally
low $^{13}$CO/C$^{18}$O line ratio.  Abundances anomalies may,
therefore, explain the globally low CO/C$^{18}$O and
$^{13}$CO/C$^{18}$O ratios, but not their spatial variation.

In summary, from the close connection between $R_{13}$ and HCN
intensities combined with the overall inadequacies of all other
explanations, it appears that the most likely explanation for the
distribution of large $R_{13}$ and $R_{18}$ gas in the nucleus of NGC
6946 is a significant and coherent decrease in gas density as one
moves away from the central ring, causing the isotopomers to become
subthermal and faint relative to the main species, which remains
thermal, opaque, and bright.  Operating in tandem with higher
temperatures or anomalous abundances that are peculiar to the nuclear
environment, a small reduction in opacity at low gas densities can
increase $R_{13}$ substantially.  The molecular arms in the inner 300
pc region appear to consist of a warm, low density gas that abruptly
transitions to warmer, dense gas in the central ring.  So the observed
$R_{13}$ is likely best considered a probe of the mixing fraction of
dense, low $R_{13}$ gas and diffuse, high R$_{13}$ gas
\citep[][]{ABBJ95,SHSHS97}.  HCN(1-0) emission probes the hotter, high
density cores embedded inside the diffuse medium.  Since there is no
global correlation between $R_{13}$ and HCN(3-2)/HCN(1-0)
\citep[][]{Pet01}, a tracer of the density of the dense gas component,
$R_{13}$ does not appear strongly sensitive to the exact density of
the dense component.
 
\subsection{High $\bf R_{13}$: Bars as Redistributors of Gas Physical 
Conditions?}

\begin{figure}
\epsscale{0.7}
\plotone{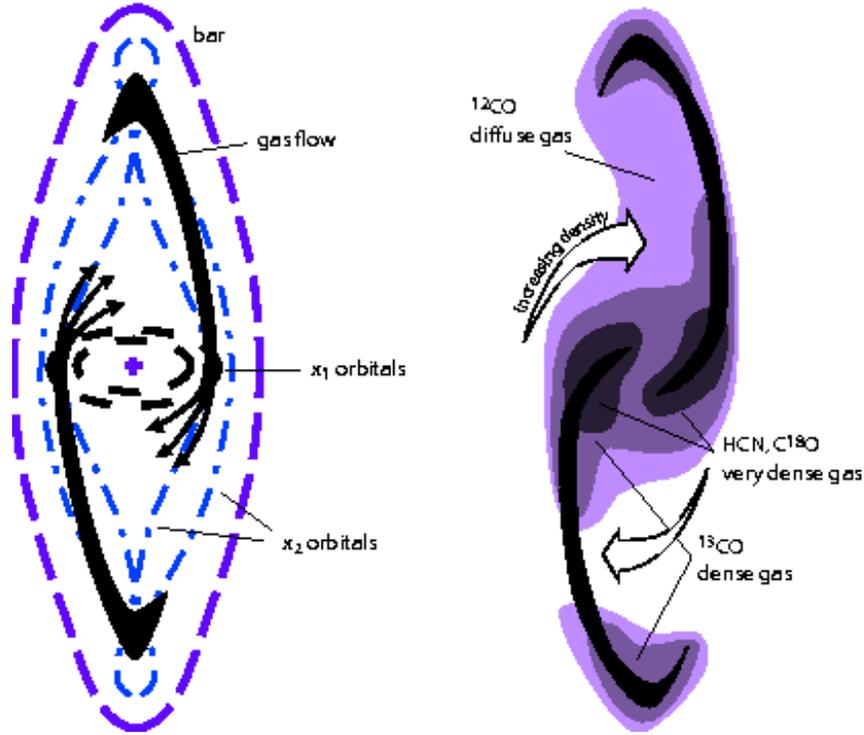}
\caption{{\it Left)} The basic flow patterns of molecular gas in a
barred potential \citep[eg.,][]{A92}.  {\it Right)} A schematic
showing the distribution of components with different physical
conditions in response to that barred potential.  The densest gas,
bright primarily in HCN and C$^{18}$O emission is confined to the
intersection regions of the $x_{1}$ and $x_{2}$ regions, explaining
the sort of ``saddle'' region seen in $R_{13}$ and CO/HCN
\citep[][]{HB97}.  $^{13}$CO is seen primarily towards the central
ring-like region as well as at the cuspy ends of the bar.  CO, on the
other hand, is remains bright over much of the entire region, tracing
the distribution of the low density gas.  The locations of the
extended, low density gas component inferred from the regions of high
$R_{13}$ correspond to the theoretically predicted ``spray regions''
in barred potentials. }
\label{galn6schem} 
\end{figure}

We find that the two gas components with distinctly different
properties ($n_{H_2}$ and \tk ) that are required to explain the
single-dish line ratios in NGC~6946 are also found in distinctly
different parts of the nucleus.  Molecular gas with high $R_{13}$
tends to be located along and upstream of the arms.  The correlation
of gas properties with position along a molecular bar was first noted
by \citet[][]{WITHL93}, in IC 342.  Similar features are now seen here
in NGC 6946, Maffei 2 (Meier, Turner \& Hurt 2004, in preparation), as
well as the much larger-scale bars, UGC 2855 and NGC 7479
\citep[][]{HAW99,HADW00}.  The morphology of $R_{13}$ in barred
galaxies such as NGC~6946 suggests that changes in $R_{13}$ are
connected to the dynamics but not through dynamical changes in
linewidth.  It must result instead from a mechanism capable of
dispersing the molecular gas into spatially distinct regions of low
density, diffuse emission and high density cloud cores, correlated
with position across the bar.  The theoretically predicted ``spray
regions'' in barred potentials \citep[][]{A92} provide a natural
explanation for the presence and locations of the extended, low
density gas component inferred from the regions of high $R_{13}$.
Dense gas is associated with the central $x_{2}$-type orbits and the
cuspy ends of the $x_{1}$-type orbits, while the remaining molecular
gas is ``scattered'' off the $x_{2}$-orbitals, putting the diffuse gas
at and upstream of the molecular arms (as illustrated in Figure
\ref{galn6schem}).  In the case of IC 342, which has a higher column
density (and density) in the arms, $^{13}$CO and C$^{18}$O extend
further along the arms than in NGC 6946.  In fact, in IC 342, the
densities are such that one can trace the density progression across
the molecular arm from the upstream side up to the (shocked) leading
edge \citep[][]{WITHL93,MT01}.  Evidently, bars acting to disperse
molecular gas into distinct regions with different physical conditions
may be a common phenomena.  This dynamical interpretation further
supports the notion that the large $R_{13}$ seen in distant mergers
reflect changes in gas physical conditions associated with the
large-scale dynamical redistribution of molecular gas.  Though in this
case, the redistribution would not be from a barred potential but from
the action of the merger, directly.
 
\subsection{High $\bf R_{13}$: Comparisons to Distant Galaxies and ULIRGs}

The resolved structure of NGC 6946 appears to be similar, but more
modest version of the structure inferred for the ISM of ULIRGs
\citep[eg.,][]{SDRB97,DS98}.  The diffuse arms in NGC 6946 appear to
be cooler, lower column density analogues of the low density, high
filling factor CO gas common in ULIRGs, while the starburst regions
that are bright in HCN and $^{13}$CO in NGC 6946 are the more modest
counterparts of the ``extreme starbursts'' seen in ULIRGs
\citep[][]{DS98}.

The ratio of CO to HCN luminosity is $\sim$4-7 in ULIRGs as compared
to $\sim$30-70 in the disk of our Galaxy and normal galaxies
\citep[eg.,][]{SDR92,SNKN02}.  The low absolute CO/HCN ratio in ULIRGs
argues strongly that the fraction of the mass in the dense component
is much higher than in more modest starbursts like NGC 6946
\citep[][]{SDR92}.  However, this exacerbates the problem of
explaining the high observed $R_{13}$.  Just requiring a large filling
factor, diffuse component along with low filling factor embedded dense
cores, like seen in NGC 6946, is not enough to simultaneously explain
both the low CO/HCN and high $R_{13}$ ratios in ULIRGs, for any
reasonable [CO/$^{13}$CO] abundance.  Evidently, one must also require
that $^{13}$CO/HCN be significantly lower than the unity observed for
NGC 6946's dense component.  This can be achieved by a large opacity
in the HCN(1-0) transition.

The relative opacity of lines with similar frequencies and upper
energy levels (as is the case for the J=1-0 transitions of CO,
$^{13}$CO, C$^{18}$O and HCN) depend primarily on their relative
abundances and transition line strengths (proportional to the dipole
moment squared, $|\mu|^{2}$).  In gas with densities high enough to
thermalize HCN (and consequently CO and $^{13}$CO),
\citep[eg.][]{KKV99}:
\begin{equation} 
\frac{\tau_{HCN}}{\tau_{13CO}} ~\simeq ~  
\left[\frac{HCN}{^{13}CO}\right] 
\left(\frac{\mu_{HCN}}{\mu_{13CO}}\right)^{2} ~\sim ~ 10, 
\end{equation} 
given expected HCN abundances.  Hence, the opacity of HCN can
significantly exceed the opacity of $^{13}$CO in warm, dense regions
\citep[eg.,][]{ARSS97}.  For example, based on our LVG models (\S4.1),
for T$_{k}$ = 150 K, $n_{H_{2}}$ = 10$^{5}$, [$^{13}$CO/$\Delta v$] =
$10^{-6.4}$ and [HCN/$\Delta v$] = $10^{-8.3}$,
$^{13}$CO(1-0)/HCN(1-0) $\simeq$ 0.3.  This value can be decreased
even more if one lowers the [CO/$^{13}$CO] abundance below that
adopted for NGC 6946 (stronger isotope-selective photodissociation?).
For the localized dense cores, CO and HCN can be bright due to high
opacity, while the optically thin $^{13}$CO remains relatively faint.
Assuming these dense cores are embedded in a low density medium with a
much larger filling factor, then it becomes possible to simultaneously
explain low beam averaged CO/HCN and CO(2-1)/CO(1-0) line ratios while
maintaining high beam-averaged $R_{13}$ ratios.  Given the fairly high
excitation required, it makes sense that this would be seen primarily
in the ``extreme starburst'' cores associated with ULIRGs.

\section{Conclusions} 
 
Aperture synthesis maps of $^{13}$CO and C$^{18}$O in the nucleus of
NGC 6946 are presented.  The gas morphology and kinematics are
consistent with molecular gas moving in response to a barred
potential.  The emission from \thco\ and \ceo\ is more symmetric and
confined to the central ring of molecular material than is CO.
Because of this, $R_{13}$ (ratio of integrated intensity,
$^{12}$CO/$^{13}$CO, J=1-0 for both) and $R_{18}$
($^{12}$CO/$^{18}$CO, J=1-0) increase dramatically moving away from
the central starburst, particularly along and to the trailing side of
the CO bright northern molecular arm.  $R_{13}$ reaches measured
values as high as 30, more than three times higher than Galactic
values, and upper limits significantly higher yet in some parts of
this region.  The transition between low $R_{13}$ gas and high
$R_{13}$ gas is very sharp once the central ring is left, indicating a
very rapid change in physical conditions.
 
If $^{13}$CO and C$^{18}$O accurately trace the amount of molecular
gas present, the standard conversion factor, $X_{CO}$, in NGC 6946
must be lower than the Galactic value.  Column densities inferred from
$^{13}$CO, C$^{18}$O, and 1.4 mm dust continuum emission imply that
$X_{CO}(6946)$ is $\sim$ 1/3 - 1/5 the Galactic value at the sites of
the GMCs, for a typical value of $\rm X_{CO}=0.5 \times
10^{20}~cm^{-2}(K ~km~ s^{-1})^{-1}$.  Towards the diffuse gas
component along the northern arm, the conversion factor is likely a
factor of 2 - 3 lower yet.  Star formation as traced by 2.7 mm
continuum and P$\alpha$ more closely follows total gas surface density
(or perhaps density) traced by the isotopomers rather than the
optically thick CO intensity. The total gas mass in the central 175
kpc radius of NGC 6946 is $\sim 1\times 10^8~M_\odot$, and the star
formation efficiency on 100 pc sizescales is a few percent.
 
Both gas moving with large radial velocities ($v_{r} \simeq 80$ km
s$^{-1}$) and bright HCN(1-0) emission imply that molecular gas has
piled up along the central ring.  LVG models find best fit densities
of $n_{H_{2}} \sim 10^{4.2}$ cm$^{-3}$ towards the central molecular
ring and densities below $n_{H_{2}} \lsim 10^{2}$ cm$^{-3}$ along the
northern arm.  Corresponding kinetic temperatures are T$_{k}\sim$90 K
for the central starburst and T$_{k}$ $\lsim$ 40 K for northern arm.
The morphology of the isotopomers are very similar to the high
critical density tracer HCN(1-0), demonstrating that density changes
are a key physical condition governing the local changes in $R_{13}$.
Together with the fact that $R_{13}$ is correlated with position
relative to the bar, this suggests that the bar potential acts to
disperse the molecular gas into coherent regions with different
physical conditions. Here in NGC~6946, the high $R_{13}$ line ratio
values are a signpost of a dynamically evolving ISM.  Applying these
findings regarding the ISM structure of NGC 6946 to other more distant
starbursts can successfully explain some of their observed molecular
properties, hinting that similar processes may be at work there too.
 
\acknowledgements 

We are grateful to the faculty and staff at OVRO for their support and
assistance during the observations.  We thank Eva Schinnerer for
helpful discussions and S. Ishizuki for the use of his $^{12}$CO(1-0)
data on NGC 6946.  We also thank an anonymous referee for providing a
critical reading of the manuscript.  This work is supported in part by
NSF grant AST-0071276 to JLT.  DSM acknowledges support from NSF grant
1-5-29627 to the Laboratory of Astronomical Imaging at the University
of Illinois.  The Owens Valley Millimeter Interferometer is operated
by Caltech with support from the NSF under Grant 9981546.

\clearpage 
 
\clearpage

\begin{deluxetable}{lcc} 
\tablenum{1} 
\tablewidth{0pt} 
\tablecaption{NGC 6946 Basic Data} 
\tablehead{ 
\colhead{Characteristic}   
&\colhead{NGC 6946}  
&\colhead{Reference} \\ 
} 
\startdata 
Hubble Class   &SAB(rs)cd   & 1  \\ 
Near-IR Center: & & \\  
$ \alpha(J2000)$ & $20^{h}34^{m}52^{s}.34\pm0.2^{''}$  &2 \\ 
$ \delta(J2000)$ &$+60^{o}09^{'}13.^{''}7\pm0.2^{''}$  &   \\ 
Radio Continuum Peak: & & \\  
$ \alpha(J2000)$ & $20^{h}34^{m}52^{s}.20\pm0.2^{''}$  &3 \\ 
$ \delta(J2000)$ &$+60^{o}09^{'}14.^{''}2\pm0.2^{''}$  &   \\ 
Dynamical Center: & & \\  
$ \alpha(J2000)$ & $20^{h}34^{m}52^{s}.29\pm2^{''}$  &4 \\ 
$ \delta(J2000)$ &$+60^{o}09^{'}14.^{''}5\pm2^{''}$  &   \\ 
V$_{lsr}$  & 50 km s$^{-1}$ &4   \\ 
Adopted Distance  & 5.2 Mpc & 5   \\ 
$i$  & 40$^{o}$  & 6  \\ 
P.A.  &242$^{o}$  & 6  \\ 
L$_{IR}$ & $1.5\times 10^{9}$ L$_{\odot}$ & 7 \\ 
T$_{d}$ &40 K & 7 \\ 
\enddata 
\tablerefs{(1) RC2 
(2) measured from 2MASS images 
(3) Turner \& Ho 1983, 
Tsai et al. 2004, in preparation, 2 cm peak;  
(4) This paper;  
(5) Tully 1988; (6) Crosthwaite \& Turner 2004, in preparation;  
(7) Engargiola 1991}  
\label{galbasdat} 
\end{deluxetable} 
 
\clearpage 
 
\begin{deluxetable}{lcccccc} 
\tablenum{2} 
\tablewidth{0pt} 
\tablecaption{Observational Data\tablenotemark{a}} 
\tablehead{ 
\colhead{Transition}  
&\colhead{$\nu_{o}$}  
&\colhead{T$_{sys}$}  
&\colhead{$\Delta V_{chan}$}  
&\colhead{$\Delta \nu_{band}$}  
&\colhead{Beam}  
&\colhead{Noise level} \\ 
\colhead{} 
&\colhead{(GHz)}  
&\colhead{(K)}  
&\colhead{(km s$^{-1}$)}  
&\colhead{(MHz)}  
&\colhead{($^{''}$;deg)}  
&\colhead{(mK/mJy Bm$^{-1}$)} \\ 
} 
\startdata 
 $^{13}$CO(1-0)& 110.20 &240-380 &10.88 &128  
&5.4x4.2;-82 & 44/10\\ 
 C$^{18}$O(1-0)& 109.78 & 240-380 & 10.92& 128  
&5.5x4.4;-83 & 44/10 \\ 
 2.7 mm\tablenotemark{b} & 110.0   &240-380 &\nodata  &1000  
&5.5x4.3;-89 & 3.8/0.9\\ 
 1.4 mm\tablenotemark{c} & 221.9   &700-1600 &\nodata  &1000  
&5.5x4.3;-89 & 3.7/5.5\\ 
\enddata 
\tablenotetext{a}{Dates for observations of NGC 6946 
were 1999 September 26 - 1999 November 28, with a phase center: 
V$_{LSR}$=50 $km ~s^{-1}$ $\alpha=20^{h}33^{m}49^{s}.20$, 
$\delta=+59^{o}58'49.^{''}0$ (B1950).} 
\tablenotetext{b}{$^{13}$CO(1-0) line contamination has been  
removed.} 
\tablenotetext{c}{Convolved to the resolution of 2.7 mm.} 
\label{galt2} 
\end{deluxetable} 
 
\clearpage 
 
\begin{deluxetable}{lccccc} 
\tablenum{3} 
\tablewidth{0pt} 
\tablecaption{Measured Intensities \& Temperatures in NGC 6946\tablenotemark{a}} 
\tablehead{ 
\colhead{GMC\tablenotemark{b}}  
&\colhead{I($^{12}$CO(1-0))\tablenotemark{c}}  
&\colhead{I($^{13}$CO(1-0))}  
&\colhead{T($^{13}$CO(1-0))}  
&\colhead{I(C$^{18}$O(1-0))}   
&\colhead{T(C$^{18}$O(1-0))} \\ 
\colhead{} 
&\colhead{($K~km~s^{-1}$)}  
&\colhead{($K~km~s^{-1}$)}  
&\colhead{($K$)}  
&\colhead{($K~km~s^{-1}$)}  
&\colhead{($K$)} \\ 
} 
\startdata 
A & 100$\pm$16 & 6.4$\pm$2 & 0.28$\pm$0.05 & $<$3.6     & 0.13$\pm$0.05\\ 
B & 120$\pm$16 & 17$\pm$2  & 0.59$\pm$0.06 & 9.0$\pm$1.8& 0.17$\pm$0.05\\ 
C & 87$\pm$16  & 6.4$\pm$2 & 0.29$\pm$0.05 & 7.2$\pm$1.8& 0.15$\pm$0.05\\ 
D & 74$\pm$16  & 8.6$\pm$2 & 0.50$\pm$0.05 & 11$\pm$1.8 & 0.20$\pm$0.05\\ 
E1& 650$\pm$65 & 64$\pm$6  & 0.82$\pm$0.08 & 31$\pm$3   & 0.33$\pm$0.05\\ 
E2& 590$\pm$59 & 50$\pm$5  & 0.59$\pm$0.06 & 10$\pm$1.8 & 0.22$\pm$0.05\\ 
F & 300$\pm$30 & 20$\pm$2  & 0.32$\pm$0.05 & 9.4$\pm$1.8& $\sim$0.11   \\ 
G & 300$\pm$30 & 20$\pm$2  & 0.45$\pm$0.05 & $<$3.6     & 0.22$\pm$0.05\\ 
H & 74$\pm$16  & 3.6$\pm$2 & 0.19$\pm$0.05 & $<$3.6     &$<$0.11    \\ 
ARM\tablenotemark{d} &270$\pm$27& 4.7$\pm$2 & 0.14$\pm$0.05  
& $<$3.6   &$<$0.11\\ 
\enddata 
\tablenotetext{a}{Uncertainties are the larger of the  
map uncertainty or the absolute calibration uncertainties.  Upper  
limits are 2$\sigma$.} 
\tablenotetext{b}{Positions are those labeled in Table \ref{galgmc} 
and Figure \ref{galn6pv}a.}
\tablenotetext{c}{Taken from the Sakamoto et al. 1999 data.} 
\tablenotetext{d}{The position of the 'ARM' measurement corresponds  
to $ \alpha(J2000)~=~20^{h}34^{m}51^{s}.20$, $\delta(J2000)~=~  
+60^{o}09^{'}28.^{''}4$.} 
\label{galn6itab} 
\end{deluxetable} 
 
\clearpage 
 
\begin{deluxetable}{ccccccc} 
\tablenum{4} 
\tablewidth{0pt} 
\tablecaption{Giant Molecular Clouds in NGC 6946\tablenotemark{a}} 
\tablehead{ 
\colhead{GMC}  
&\colhead{$\alpha, ~\delta$} 
&\colhead{$a \times b;~ pa$} 
&\colhead{$\Delta v_{1/2}$} 
&\colhead{$v_{o}$} 
&\colhead{$M_{vir}$\tablenotemark{b}} 
&\colhead{$^{13}M_{CO}$\tablenotemark{c}}  \\  
\colhead{} 
&\colhead{(20$^{h}$34$^{m}$;60$^{o}$09$^{'}$)}  
&\colhead{($pc \times pc;^{o}$)}  
&\colhead{($km~s^{-1}$)} 
&\colhead{($km~s^{-1}$)} 
&\colhead{($10^{6}~M_{\odot}$)} 
&\colhead{($10^{6}~M_{\odot}$)} \\ 
} 
\startdata 
A  &53.56;53.1\tablenotemark{d} &230x68;44  & 41& 55 
&28 &1.3  \\ 
B  &53.44;19.0 &260x150;149  & 26& -19 & 18 &8.6 \\ 
C  &52.68;59.6\tablenotemark{d} &110x$<$53\tablenotemark{e}  & 40  
& 104 &  $<$16 &1.2  \\ 
D &52.85;51.4\tablenotemark{d} &290x140;90  & 35 & 91 & 33 & 4.4 \\ 
E1 &52.62;13.8 &170x100;137  & 110& 48 &210 &14   \\ 
E2 &51.93;15.9 &270x$<$53;148\tablenotemark{e} & 100& 83 &  $<$160 & 9.2\\ 
F  &51.74;22.9 &110x68;28  & 97 & 37 &  110 &3.9 \\ 
G  &51.52;40.0 &180x140;68  & 46& 62 & 44 &6.6 \\ 
H  &50.72;05.9 &150x66;35  & 22 & 133 &6.4 & 0.66 \\ 
\enddata 
\tablenotetext{a}{Typical fit uncertainties are less  
1/2 the channel width of 10.9 K km s$^{-1}$, so an uncertainty of  
5.5 K km s$^{-1}$ is conservatively estimated.} 
\tablenotetext{b}{Based on 189$(\Delta v_{1/2;km/s})^{2}R_{pc}$, with  
R defined as 1.4R$_{FWHM}$, or 0.7$\sqrt{ab}$ (eg. Meier \& Turner  
2001).} 
\tablenotetext{c}{Based on the molecular column density from  
$^{13}$CO(1-0) corrected for resolved out emission and He content,  
and the larger of the beamsize or fitted cloud size.} 
\tablenotetext{d}{Based on 20$^{h}$34$^{m}$; 60$^{o}$08$^{'}$.}  
\tablenotetext{e}{A GMC is considered unresolved if its  
deconvolved size is less than 1/2 of the beam minor axis.} 
\label{galgmc} 
\end{deluxetable} 
 
\clearpage

\begin{deluxetable}{lccc} 
\tablenum{5} 
\tablewidth{0pt} 
\tablecaption{Observed Line Ratios in NGC 6946} 
\tablehead{ 
\colhead{GMC} 
&\colhead{$\frac{^{12}CO(1-0)}{^{13}CO(1-0)}$}   
&\colhead{$\frac{^{12}CO(1-0)}{C^{18}O(1-0)}$} 
&\colhead{$\frac{^{13}CO(1-0)}{C^{18}O(1-0)}$} \\ 
} 
\startdata 
A & 16$\pm$6  & $>$28      & $>$1.8 \\  
B & 7.1$\pm$1 & 13$\pm$3   & 1.9$\pm$0.4 \\  
C & 14$\pm$5  & 12$\pm$5   & $\sim$0.9 \\  
D & 8.6$\pm$3 & 6.7$\pm$2  & $\sim$0.8 \\  
E1& 10$\pm$2  & 21$\pm$3   & 2.1$\pm$0.3 \\  
E2& 12$\pm$2  & 59$\pm$12  & 5.0$\pm$1 \\  
F & 15$\pm$2  & 32$\pm$7   & 2.1$\pm$0.5 \\  
G & 15$\pm$2  & $>$170     & $>$11 \\  
H & 21$\pm$13 & $>$21      & $>$1    \\  
ARM & 57$\pm$25 & $>$75      & $>$1    \\  
\enddata 
\label{galn6rattab} 
\end{deluxetable} 
 
\clearpage 
 
\begin{deluxetable}{lcccccc} 
\tablenum{6} 
\tablewidth{0pt} 
\tablecaption{Column Densities in NGC 6946$^{a}$} 
\tablehead{ 
\colhead{GMC} 
&\colhead{$^{Xco}N_{H_{2}}$} 
&\colhead{$^{13}N_{H_{2}}$\tablenotemark{b}} 
&\colhead{$^{18}N_{H_{2}}$\tablenotemark{c}} 
&\colhead{$^{V}N_{H_{2}}$\tablenotemark{d}} 
&\colhead{$\frac{^{MW}X_{CO}}{^{13}X_{co}}$} 
&\colhead{$\frac{^{MW}X_{CO}}{^{18}X_{co}}$}\\ 
\colhead{} 
&\colhead{($10^{22}cm^{-2}$)} 
&\colhead{($10^{22}cm^{-2}$)} 
&\colhead{($10^{22}cm^{-2}$)} 
&\colhead{($10^{22}cm^{-2}$)} 
&\colhead{} 
&\colhead{} \\ 
} 
\startdata 
A & 2.0$\pm$0.3 & 0.31$\pm$0.1 & $<$0.37     & 6.8 & 6.7$\pm$2  
& $>$5.4 \\ 
B & 2.4$\pm$0.3 & 0.83$\pm$0.1 & 0.96$\pm$0.2 &\nodata & 2.9$\pm$0.5  
& 2.5$\pm$0.7 \\ 
C & 1.7$\pm$0.3 & 0.31$\pm$0.1 & 0.72$\pm$0.2 &$<$4.0 & 5.5$\pm$2  
& 2.4$\pm$0.7 \\ 
D & 1.5$\pm$0.3 & 0.41$\pm$0.1 & 1.1$\pm$0.2 & 3.1& 3.7$\pm$1  
& 1.4$\pm$0.3 \\ 
E1 & 13$\pm$1.3 & 3.1$\pm$0.3  & 3.2$\pm$0.2 &\nodata & 4.2$\pm$0.7  
& 4.1$\pm$0.7 \\ 
E2 & 12$\pm$1.2 & 2.4$\pm$0.3  & 1.0$\pm$0.2 &\nodata & 5.0$\pm$0.8  
& 12$\pm$3 \\ 
F & 6.0$\pm$0.6 & 0.98$\pm$0.1 & 0.96$\pm$0.2 &\nodata & 6.1$\pm$1  
& 6.3$\pm$1 \\ 
G & 6.0$\pm$0.6 & 0.98$\pm$0.1 & $<$0.37     &6.6 & 6.1$\pm$1  
& $>$16 \\ 
H & 1.5$\pm$0.3 & 0.17$\pm$0.1 & $<$0.37     &1.6 & 8.8$\pm$5  
& $>$4.1 \\ 
ARM\tablenotemark{e} & 5.4$\pm$0.5 & 0.22$\pm$0.1 & $<$0.37      
&\nodata & 25$\pm$10 & $>$15 \\ 
\enddata 
\tablenotetext{a}{None of the column densities have not 
been corrected for resolved-out flux.  To account for this 
$^{13}N_{H_{2}}$ (and presumably the $^{18}N_{H_{2}}$) should be 
divided by $\sim 0.63$ and $^{Xco}N_{H_{2}}$ should be divided by 
$\sim 0.76$.  Uncertainties reflect only the statistical uncertainties 
and not the potentially larger systematic uncertainties associated 
correct abundance ratio and T$_{ex}$ (see text).  Upper Limits are  
2$\sigma$.} 
\tablenotetext{b}{Based on the LTE assumption with  
T$_{ex}$ = 20 K, and [$^{12}$CO/$^{13}$CO]= 40.} 
\tablenotetext{c}{Based on the LTE assumption with  
T$_{ex}$ = 10 K, and [$^{12}$CO/C$^{18}$O]= 150.}  
\tablenotetext{d}{Based on the virial masses derived for the  
isolated GMCs (Table \ref{galgmc}).} 
\tablenotetext{e}{Based on the position for 'ARM' given in Table  
\ref{galn6itab}.} 
\label{galn6col} 
\end{deluxetable}

\end{document}